\pdfoutput=1
\documentclass[a4paper,11pt]{article}
\usepackage{amsmath,amssymb,amsfonts}
\usepackage[multiple,stable]{footmisc}
\usepackage[amsmath,amsthm,thmmarks]{ntheorem}
\allowdisplaybreaks[4]
\usepackage[noconfig]{refstyle}
\usepackage[dvipsnames]{xcolor}
\usepackage[hyperfootnotes=false, linktocpage=true, colorlinks, citecolor=blue, linkcolor=blue, urlcolor=Maroon]{hyperref}
\usepackage{array,tabularx,booktabs,geometry,subfigure,graphicx,longtable}
\usepackage[bf]{caption}
\usepackage{tikz}
\usepackage{cite}
\usetikzlibrary{shapes,arrows}
\tikzstyle{block} = [rectangle, draw, text width=7em, text centered, rounded corners, minimum height=3em]

\let\eqref=\relax
\newref{eq}{name={eq.~},Name={Eq.~},names={eqs.~},Names={Eqs.~},rngtxt={-},refcmd=(\ref{#1})}
\newref{f}{name={footnote~},Name={Footnote~},names={footnotes~},Names={Footnotes~}}
\newref{app}{name={appendix~},Name={Appendix~},names={appendixes~},Names={Appendixes~}}
\newref{tab}{name={table~},Name={Table~},names={tables~},Names={Tables~}}
\newref{ch}{name={chapter~},Name={Chapter~},names={chapters~},Names={Chapters~}}
\newref{sec}{name={section~},Name={Section~},names={sections~},Names={Sections~}}
\newref{fig}{name={figure~},Name={Figure~},names={figures~},Names={Figures~}}
\numberwithin{equation}{section}

\newcommand{\be}{\begin{equation}}
\newcommand{\ee}{\end{equation}}
\newcommand{\bea}{\begin{equation}\begin{aligned}}	
\newcommand{\eea}{\end{aligned}\end{equation}}		

\newcommand{\tr}{\mathrm{tr}}

\newcommand{\iddots}{\mathinner{\mkern2mu\raise1pt\hbox{.}\mkern2mu \raise4pt\hbox{.}\mkern2mu\raise7pt\hbox{.}\mkern1mu}}

\providecommand{\id}{\leavevmode\hbox{\small$\mathrm{1}$\kern-3.8pt\normalsize$\mathrm{1}$}}
\def\fnote#1#2{\begingroup\def\thefootnote{#1}\footnote{#2}
     \addtocounter{footnote}{-1}\endgroup}

\begin{document}

\vspace{1cm}

\title{       {\Large \bf  Moduli identification methods in Type II compactifications}}

\vspace{2cm}

\author{
James~Gray${}{}$ and
Hadi~Parsian${}{}$
}
\date{}
\maketitle
\begin{center} {\small ${}${\it Department of Physics, 
Robeson Hall, Virginia Tech \\ Blacksburg, VA 24061, U.S.A.}}

\fnote{}{jamesgray@vt.edu}
\fnote{}{varzi61@vt.edu }

\end{center}

\begin{abstract}
\noindent
Recent work on four dimensional effective descriptions of the heterotic string has identified the moduli of such systems as being given by kernels of maps between ordinary Dolbeault cohomology groups. The maps involved are defined by the supergravity data of the background solutions. Such structure is seen both in the case of Calabi-Yau compactifications with non-trivial constraints on moduli arising from the gauge bundle and in the case of some non-K\"ahler compactifications of the theory. This description of the moduli has allowed the explicit computation of the moduli stabilization effects of a wide range of non-trivial gauge bundles on Calabi-Yau three-folds. In this paper we examine to what extent the ideas and techniques used in this work can be extended to the case of Type IIB string theory. Certain simplifications arise in the Type IIB case in comparison to the heterotic situation. However, complications also arise due to the richer supergravity data of the theory inducing a more involved map structure. We illustrate our discussion with several concrete examples of compactification of Type IIB string theory on conformal CICY three-folds with flux.

\end{abstract}

\thispagestyle{empty}
\setcounter{page}{0}
\newpage

\tableofcontents

\section{Introduction}

In a series of recent work, it has been shown that the moduli of compactifications of the heterotic string can frequently be written as sums of kernels of maps between Dolbeault cohomology groups \cite{Anderson:2010mh,Anderson:2011cza,Anderson:2011ty,Anderson:2013qca,Anderson:2014xha}. Cases studied include non-trivial slope zero poly-stable holomorphic vector bundles over Calabi-Yau threefolds, and more general non-K\"ahler solutions satisying the $\partial \overline{\partial}$-lemma. The kernels in question typically live inside the naive cohomology groups one would associate with the massless degrees of freedom of a Calabi-Yau compactification, $H^1(TX)$, $H^1(TX^{\vee})$ and $H^1(\textnormal{End}_0(V))$. The maps are determined by supergravity field strengths of different types. Thus, more complicated sets of fluxes, such as those seen in non-K\"ahler compactifications, lead to a more complicated series of maps \cite{Anderson:2014xha} (see \cite{Melnikov:2011ez,delaOssa:2014cia} for related work).

The procedure to derive such a description of the massless degrees of freedom of a theory is rather straightforward. Massless moduli are in one-to-one correspondence with linear fluctuations of the fields which satisfy the higher dimensional equations of motion. Thus, one can simply perform such a perturbation and then see if the resulting `allowed fluctuations' can be written as the kernel of maps between Dolbeault cohomology groups.

Practically there are great advantages to writing the massless degrees of freedom of a theory in this language. At least in the case of vector bundles over Calabi-Yau three-folds, many examples have been given where the relevant kernels can be computed explicitly \cite{Anderson:2010mh,Anderson:2011cza,Anderson:2011ty,Anderson:2013qca} (a dearth of background solutions still plagues the subject of non-K\"ahler compactifications despite interesting recent work \cite{Goldstein:2002pg,Fu:2006vj,Fei:2015kua,Fei:2017ctw}). Indeed, the allowed complex structure fluctuations can be computed in a sufficiently concrete fashion that the restriction on the coefficients in a set of polynomial defining equations can be described precisely in complete intersection examples. Such concreteness of description makes it possible to go further and ask about the effect of stabilization on questions such as the matter spectrum, which is itself determined  by complex structure dependent Dolbeault cohomology groups. Most of the examples that have been provided so far in this work have been couched in the language of complete intersections in products of projective spaces \cite{Hubsch:1992nu}, although generalizations to other constructions such as toric hypersurfaces would be straightforward.

\vspace{0.1cm}

In this paper, we wish to address the question of how much of the technology that has been developed in the heterotic literature, in particular with regard to computing moduli explicitly in examples, can be directly applied to the case of compactifications of type IIB string theory. It should be noted that there is a large and well established literature featuring a plethora of sophisticated approaches to moduli identification and effective theory derivation in a wide variety of type IIB compactifications. We will not attempt a systematic review of this vast literature here and instead simply direct the reader to some examples of such work that are most relevant to the current paper \cite{DeWolfe:2002nn,Grana:2003ek,Grimm:2004uq,Grana:2004bg,Behrndt:2005bv,Grana:2005sn,Giddings:2005ff,Koerber:2006hh,Tomasiello:2007zq,Shiu:2008ry,Douglas:2008jx,Frey:2008xw,Marchesano:2008rg,Martucci:2009sf,Chen:2009zi,Tseng:2009gr,Underwood:2010pm,Tseng:2010kt,Marchesano:2010bs,Grana:2011nb,Coimbra:2011nw,Tseng:2011gv,Grimm:2012rg,Coimbra:2012yy,Frey:2013bha,Grana:2014vva,Marchesano:2014iea,Martucci:2014ska,Coimbra:2014uxa,Grimm:2014efa,Coimbra:2015nha,Grimm:2015mua,Carta:2016ynn,Cownden:2016hpf,Martucci:2016pzt,Sethi:2017phn}. In this work we will simply focus on those cases in which the computational power for dealing with examples, seen in the heterotic work, can be utilized.  As such, we will focus on cases where the moduli can be shown to be described as kernels of maps between ordinary Dolbeault cohomology groups, which will require us in particular to, as in the heterotic case, impose the $\partial\overline{\partial}$-lemma. After a general analysis of how this occurs, we wish to try and construct explicit examples and see what simplifications and complications arise in comparison to the heterotic case.

\vspace{0.1cm}

The rest of this paper is organized as follows. In Section \ref{het} we will review moduli identification in heterotic theories. We will begin by discussing a general analysis, valid for any compactification of a given type. We will then discuss in detail the types of calculations that have been achieved in explicit examples and the structure that has been seen therein. In Section \ref{IIB} we will perform the corresponding general analysis in the type IIB case, fluctuating the equations of motion to linear order and interpreting the resulting equations in terms of maps between Dolbeault cohomology groups. In Section \ref{IIBEGS} we will study two explicit examples of the previous general analysis, both based upon conformal Calabi-Yau compactifications. The first example will be centered around a simple freely acting quotient of the quintic Calabi-Yau threefold. The second will utilize a somewhat more complicated case in order to show that the techniques being discussed are not restricted to such trivial examples. We will discuss the differences, both positive and negative, that we find between the type IIB and heterotic string theory cases. Finally, we conclude our discussion in Section \ref{conc}.


\section{Review of the Heterotic Case} \label{het}

\subsection{General analysis}

Let us begin with the simplest example of the type of structure we are interested in \cite{Anderson:2010mh,Anderson:2011ty,Anderson:2013qca}. In an ${\cal N}=1$ compactification of the heterotic string to four dimensions on a Calabi-Yau threefold, the gauge connection must obey the so called Hermitian Yang-Mills equations.
\begin{eqnarray} \label{fandd}
g^{a \overline{b}} F_{a \overline{b}}=0\;\;,\;\; F_{ab}=F_{\overline{a}\overline{b}}=0
\end{eqnarray}
These supersymmetric constraints are well known to be the higher dimensional antecedents of requiring D- and F-flatness respectively in the associated four-dimensional theory \cite{Witten:1985bz,Anderson:2009sw}. In this paper we will focus on F-flatness conditions.

One can ask, in a situation where one has a supersymmetric vacuum, what are the constraints on fluctuations around that vacuum such that supersymmetry is preserved. Such fluctuations will correspond to the massless degrees of freedom of the associated four dimensional effective theory. For the case of the holomorphy constraint $F_{\overline{a}\overline{b}}=0$, these conditions are easy to compute. We must vary all of the degrees of freedom appearing in the equation: in this case the gauge field (with fluctuation $\delta  A$) and the complex structure tensor (with fluctuation $\delta {\cal J}$). The following constraints are obtained \cite{Anderson:2010mh,Anderson:2011ty}.
\begin{eqnarray} \label{thisone}
\delta {\cal J}_{[\overline{a}}^{\;d} F^{(0)}_{\overline{b}]d}+i D^{(0)}_{[\overline{a}} \delta A_{\overline{b}]} =0
\end{eqnarray}
Thus, a complex structure fluctuation $\delta {\cal J} \in H^1(TX)$ is a true low energy degree of freedom iff there exists a $\delta A$ which solves (\ref{thisone}). If this is not the case then, under such a change in complex structure of the base manifold, the bundle associated to the heterotic compactification can not adjust so as to remain holomorphic. The quantity $\delta A$, in an instance where (\ref{thisone}) has a solution, forms part of the dimensional reduction ansatz used to obtain the four dimensional heterotic effective theory.

The equation (\ref{thisone}) can be interpreted as saying that the allowed complex structure fluctuations (those which correspond to massless modes in the low energy theory) are described by the following kernel of a map between cohomology groups \cite{Anderson:2010mh,Anderson:2011ty}.
\begin{eqnarray}  \label{map1}
\ker \left( H^1(TX) \stackrel{F^{(0)}}{\longrightarrow} H^2(\textnormal{End}_0(V)) \right)
\end{eqnarray}
The map in (\ref{map1}) is defined in terms of the unperturbed field strength via the first term in equation (\ref{thisone}) which can be verified to indeed provide a well defined map in cohomology. The fact that this must be canceled by the second term, which is exact, then tells us that allowed complex fluctuations will map to the trivial cohomology class, as indicated by the kernel in (\ref{map1}).

The gauge field fluctuations are much easier to interpret in this case. The other type of solution we can have to (\ref{thisone}) is to set $\delta J=0$ and take a closed $\delta A$. After removing a redundancy due to gauge transformations, this simply states that the allowed fluctuations in the gauge connection lie in $H^1(\textnormal{End}_0(V))$, as would naively be thought.

Such a discussion is very well known in the mathematics literature and is the manifestation in effective field theory of Atiyah's analysis of the tangent space to the moduli space of holomorphic bundles. In fact the combined allowed complex structure and bundle moduli can be described as $H^1({\cal Q})$ where ${\cal Q}$ is defined by the following short exact sequence.
\begin{eqnarray} \label{atiyah1}
0 \to \textnormal{End}_0(V) \to {\cal Q} \to \textnormal{TX} \to 0
\end{eqnarray}
Taking the long exact sequence in cohomology associated to (\ref{atiyah1}), we then find the following,
\begin{eqnarray} \label{atiyah2}
H^1({\cal Q}) = H^1(\textnormal{End}_0(V)) \oplus \ker \left( H^1(TX) \stackrel{F^{(0)}}{\longrightarrow} H^2(\textnormal{End}_0(V)) \right)
\end{eqnarray}
which matches the above analysis of the allowed fluctuations.

Those degrees of freedom associated to the complex structure of the base $X$ which are removed from the massless spectrum by the kernel constraint (\ref{map1}) often obtain masses close to the compactification scale, and thus should not be considered as fields in the four dimensional effective theory. In special cases, however, these masses might be lower and in such instances we can easily see that the constraints (\ref{map1}) are simply the higher dimensional manifestation of the F-flatness condition for massless degrees of freedom in the four dimensional theory. Indeed, that this is so might be guessed from the holomorphic nature of the equation being varied.

To see this connection to F-flatness directly one can simply consider the variation of the Gukov-Vafa-Witten (GVW) superpotential \cite{Gukov:1999ya}. The heterotic superpotential is well known to contain a term of the following form.
\begin{eqnarray}  \label{superpot}
W \ni \int_X   H \wedge \Omega
\end{eqnarray} 
Here $\Omega$ is the holomorphic three form and, locally at least, the field strength appearing is given in terms of the Yang-Mills and Lorentz Chern-Simons terms, $\omega^{3YM}$ and $\omega^{3L}$, by
\begin{eqnarray} 
H= dB - \frac{3\alpha'}{\sqrt{2}} \left( \omega^{3YM} - \omega^{3L} \right)\;.
\end{eqnarray}

The scalar components of the matter fields are obtained as fluctuations in the gauge degrees of freedom. Thus, the superpotential (\ref{superpot}) depends upon the matter fields of the theory $C_i$ solely through the term including $\omega^{3YM}$. Using this information, it is easy to see that one of the conditions for a four dimensional supersymmetric Minkowski vacuum becomes the following.
\begin{eqnarray} \label{F1}
\frac{\partial W}{\partial C_i} = -\frac{3 \alpha'}{\sqrt{2}} \int_X \Omega \wedge \frac{\partial \omega^{3YM}}{\partial C_i}=0
\end{eqnarray}
Varying this supersymmetry condition with respect to both the complex structure and perturbations in the gauge field, as we did for the ten dimensional equations above, we then arrive at the following expression \cite{Anderson:2010mh}.
\begin{eqnarray} \label{F2}
\delta \left(\frac{\partial W}{\partial C_i} \right)  = \int_X \epsilon^{\overline{a}\overline{b}\overline{c}} \epsilon^{abc} \Omega_{abc}2 \overline{\omega}_{\overline{c}}^i \tr ( T_x T_y) \left( \delta {\cal J}_{[\overline{a}}^{\; d} F_{\overline{b}]d}^{(0)} + i D_{[\overline{a}}^{(0)} \delta A_{\overline{b}]} \right)
\end{eqnarray}
Here, the $T$'s are gauge generators and $\overline{\omega}^i$ is the one form associated to the matter field $C_i$. We see immediately that this F-flatness condition is satisfied, under a variation of the fields if the condition (\ref{thisone}) holds. The constraints on massless modes we have been discussing are indeed associated to F-Flatness.

The type of general analysis of the F-flat moduli space, in terms of kernels of maps between Dolbeault cohomology groups, that we have pursued above can also be carried out in the case of Non-K\"ahler compactifications of Heterotic theories. The steps in the analysis are very similar, although the resulting map structure is somewhat more involved \cite{Anderson:2014xha} (see \cite{Melnikov:2011ez,delaOssa:2014cia} for related work).

\subsection{Computing in an example}  \label{hetegs}

The above general considerations are useful in gaining an understanding of the nature of the moduli of a heterotic Calabi-Yau compactification. However, to compute more explicit details we must specialize our analysis to a given example. In particular, we must specify a Calabi-Yau threefold and a holomorphic, slope poly-stable bundle over it. 

As a simple example consider the following Calabi-Yau manifold, defined as a complete intersection in a product of projective space (or ``CICY"), and $SU(2)$ bundle, defined as an extension of two line bundles \cite{Anderson:2010mh}.

\begin{eqnarray} \label{eg1}
X= \left[ \begin{array}{c|c} \mathbb{P}^1 & 2 \\  \mathbb{P}^1&2 \\ \mathbb{P}^2 &3 \end{array}\right]^{3,75} \;\;\;\;\;,\;\;\;\;\; 0\to {\cal L} \to V \to {\cal L}^{\vee} \to 0
\end{eqnarray}
Here the line bundle ${\cal L}$ is taken to be  ${\cal L}={\cal O}_X(-2,-1,2)$ and $V$ is indeed poly-stable in appropriate regions of K\"ahler moduli space \cite{Anderson:2010mh}.

One could study the F-flat moduli space of this theory by pursuing the above approach of perturbing around a good choice of complex structure and bundle moduli. However, this would only give us a limited view into the full moduli space of the system, restricted to the neighborhood of that starting choice. In addition, guessing a suitable initial point to perturb about can be difficult in many cases.

Instead we can use the structure of bundles, such as that in (\ref{eg1}), in order to obtain a more global view of the F-flat moduli space. The non-trivial extensions we are considering here are controlled by the extension group $\textnormal{Ext}^1({\cal L}^{\vee},{\cal L})=H^1(X, {\cal L}^2)$. This cohomology group actually vanishes for a generic choice of the complex structure of $X$. Thus generically, no such holomorphic $SU(2)$ bundle exists. However, for sub-loci of complex structure moduli space, the cohomology $H^1(X, {\cal L}^2)$ can jump in dimension to a non-zero value. On such loci, one can define a non-trivial holomorphic $SU(2)$ bundle of the type desired. One may then posit that if we consider a complex structure perturbation which takes the system off of this ``jumping locus" that, because the above $SU(2)$ bundle can no longer remain holomorphic, the holomorphic restriction we studied at the start of this section would make such a degree of freedom massive. This is indeed the case as was shown in \cite{Anderson:2010mh,Anderson:2011ty}. Thus, in order to study the F-flat complex structure moduli space in such an example, we simply need to ascertain the loci where the cohomology $H^1(X, {\cal L}^2)$ jumps.

The jumping locus of a line bundle cohomology over a Calabi-Yau threefold can readily be obtained by making use of the Koszul sequence. For this codimension one example we have the following short exact sequence.
\begin{eqnarray} \label{seq1}
0 \to {\cal N}^{\vee} \otimes {\cal L}^2 \to {\cal L}^2_{\cal A} \to {\cal L}^2_X \to 0
\end{eqnarray}
Here ${\cal A}$ denotes the ambient space $\mathbb{P}^1 \times \mathbb{P}^1 \times \mathbb{P}^2$ and ${\cal N}$ the normal bundle, ${\cal O}_{\cal A}(2,2,3)$ in this case. The short exact sequence (\ref{seq1}) has an associated long exact sequence in cohomology. Using the fact that, for the ${\cal L}$ and ${\cal N}$ given above, $H^1({\cal A}, {\cal L}^2)=H^3({\cal A},{\cal N}^{\vee} \otimes{\cal L}^2) =0$ we can write the following.
\begin{eqnarray} \label{leq1}
0 \to H^1(X,{\cal L}^2) \to H^2({\cal A},{\cal N}^{\vee} \otimes {\cal L}^2) \stackrel{P}{\longrightarrow} H^2({\cal A},{\cal L}^2) \to H^2(X,{\cal L}^2) \to 0
\end{eqnarray}
Here $P$ is the map defined by the defining relation of the Calabi-Yau threefold. Using the theorem of Bott-Borel-Weil \cite{Hubsch:1992nu}, we can describe $H^2({\cal A},{\cal N}^{\vee} \otimes {\cal L}^2)=H^2({\cal A},{\cal O}(-6,-4,1))$ as the space of linear combinations of monomials of degree $\left[-4,-2,1\right]$ in the homogenous coordinates of the ambient space $\mathbb{P}^1 \times \mathbb{P}^1 \times \mathbb{P}^2$. Likewise, the space  $H^2({\cal A},{\cal L}^2)=H^2({\cal A},{\cal O}(-4,-2,4)$ can be described as the space of linear combinations of monomials of degree $\left[ -2,0,4\right]$. With these explicit descriptions of the source and target spaces of (\ref{leq1}), together with the explicit form of the map $P$, it is then easy to find a description of $H^1(X, {\cal L}^2)$ by taking the kernel of the mapping.

We describe the general element of the source by 
\begin{eqnarray}
b_i S^i \in H^2({\cal A},{\cal N}^{\vee} \otimes {\cal L}^2)
\end{eqnarray}
where the $S^i$ are a basis of monomials of the right degree and the $b_i$ are coefficients. We then multiply this general element of the source by the defining equation,
\begin{eqnarray}
P = c_a M^a \;,
\end{eqnarray}
where the $c_a$ are coefficients (actually a redundant description of the complex structure moduli space) and the $M^a$ are a basis of degree $\left[2,2,3 \right]$ monomials. We then set to zero any term in the resulting expression which is not of the degree $\left[-2,0,4\right]$ corresponding to the target space and this gives the image of the map in (\ref{leq1}). Setting to zero the coefficient of each monomial in this image then gives us the conditions on the $b_i$ and the $c_a$ for a given set of source coefficients to give rise to an element of the kernel, that is $H^1(X, {\cal L}^2)$, for a given complex structure of the base Calabi-Yau manifold. The resulting equations take the following bilinear form.
\begin{eqnarray} \label{fulleq}
\Lambda_I^{ia}b_i c_a =0
\end{eqnarray}
Here the index $I$ runs over the dimension of the target of the map and the $\Lambda$'s are simply constants. The equations (\ref{fulleq}) contain all of the information about what elements of $H^2({\cal A},{\cal N}^{\vee} \otimes {\cal L}^2)$ give rise to elements of $H^1(X,{\cal L}^2)$, and thus possible extensions classes for the bundle $V$, for all possible values of the complex structure.

\begin{figure}[!h]\centering
\includegraphics[width=0.52\textwidth]{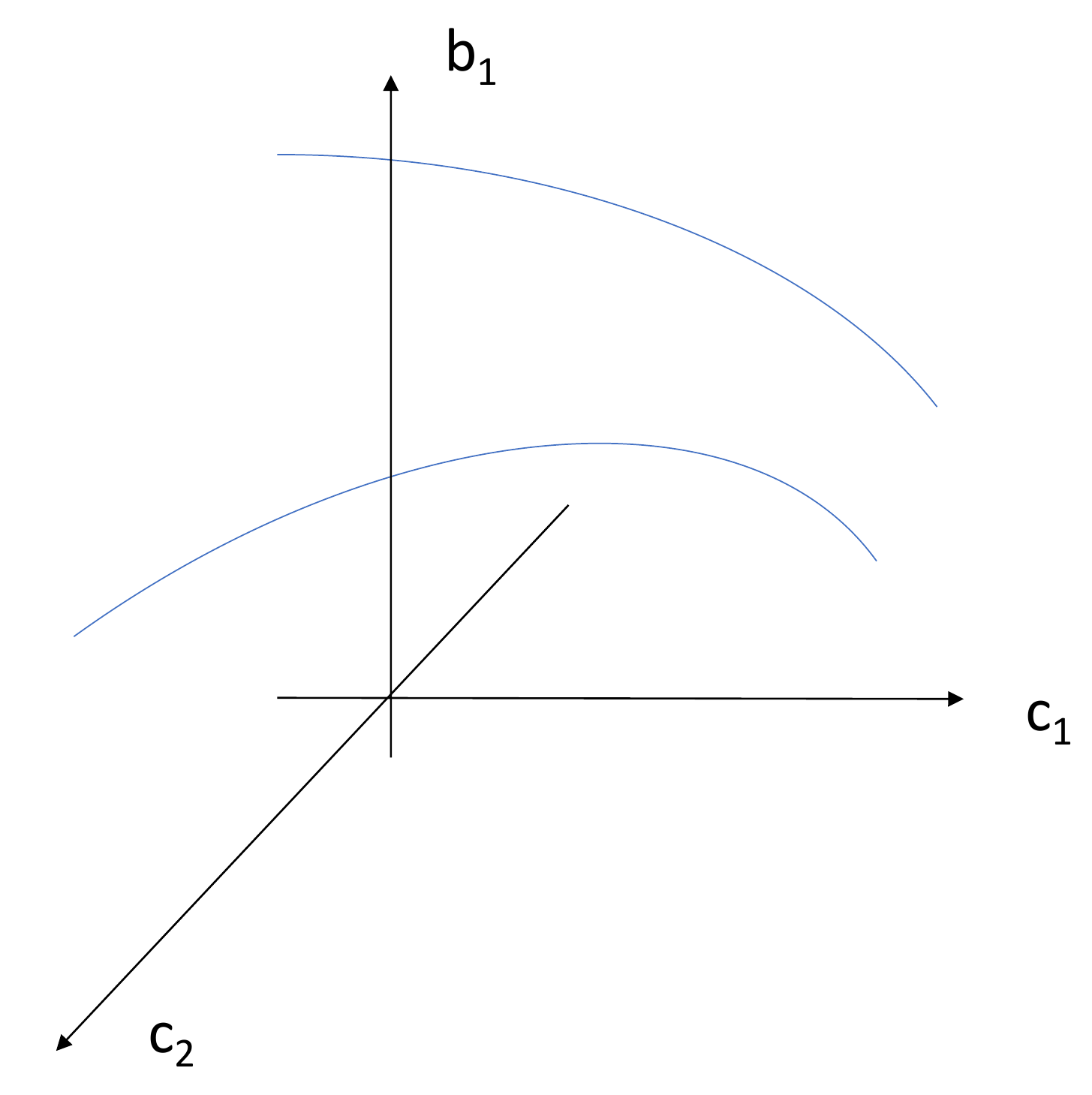}
\caption{{\it A depiction of the reducible variety given by the system (\ref{fulleq}). This variety lives in the combined space of complex structure, $c_a$, and source coefficients, $b_i$, of the map $P$ in (\ref{leq1}).}}
\label{fig1}
\end{figure}

\begin{figure}[!h]\centering
\includegraphics[width=0.89\textwidth]{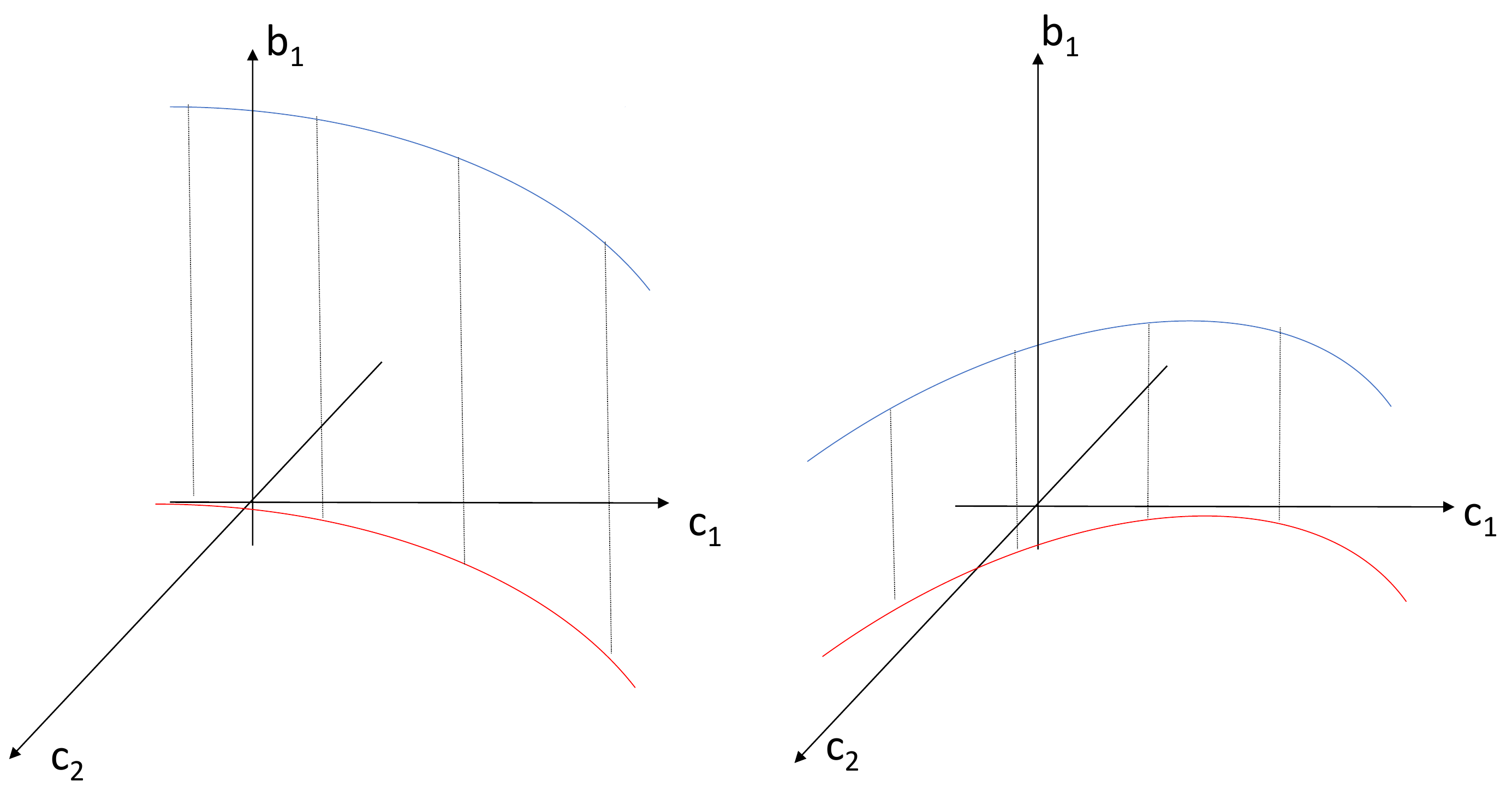}
\caption{{\it A depiction of the process of primary decomposition and then elimination as applied to the variety pictured in Figure \ref{fig1}. The red lines are the final loci obtained in complex structure moduli space.}}
\label{fig2}
\end{figure}

The set of equations (\ref{fulleq}), describes a reducible algebraic variety in the combined space of source coefficients and complex structure, as depicted in Figure \ref{fig1}. 
We can regard the source space as the space of potential elements of the kernel, with the actual elements of $H^1(X,{\cal L}^2)$ being picked out by the solutions to these equations for a given complex structure. This algebraic variety can be broken up into its irreducible components by performing a primary decomposition on the ideal whose generators are given by (\ref{fulleq}). This gives us one set of equations for each irreducible piece of the variety. By then performing an algebraic elimination of the $b$'s on each irreducible variety we can find a set of loci purely in complex structure moduli space, as parameterized by the $c$'s. This process of primary decomposition and elimination, when applied to the toy example depicted in Figure \ref{fig1}, is depicted in Figure \ref{fig2}.
These are then the loci in complex structure moduli space that the system can be stabilized to, by the effects described in the previous subsection. The system will be stabilized to a particular locus if an extension class lying in the associated set of possible $b$'s, corresponding to points on that irreducible variety, is chosen. By applying the methodology described here one can map out the whole moduli space associated to a given vector bundle, including all branches which are present due to different jumping phenomena in cohomology. For each irreducible variety in complex structure moduli space which is obtained in this manner, we must finally check that the Calabi-Yau threefold under consideration remains smooth for a generic complex structure on that locus.

In many examples, the structure of loci in complex structure moduli space to which the system can be stabilized turns out to be rather rich. For example, a simple case is given in \cite{Anderson:2013qca} for which 25 non-trivial loci are found. Of these, all but one correspond to a singular Calabi-Yau initially, although it is demonstrated that some of the rest can be smoothed out by an appropriate geometrical transition.

\vspace{0.1cm}

One thing which is important to note is that, in the above, it is vital to begin by choosing a type of bundle construction with which to work. The details of which values the complex structure get stabilized to will depend upon the structure of the bundles in question, which is not uniquely determined by topological invariants such as Chern classes.

\vspace{0.2cm}

The case of moduli stabilization due to (or more precisely moduli identification in the presence of) a holomorphic vector bundle over a Calabi-Yau threefold is just one example of how such analyses have been applied in a heterotic setting. More generally, for example, the same type of reasoning has been used to determine the F-flat moduli space of the Strominger system \cite{Strominger:1986uh}, for cases where the compactification manifold obeys the $\partial \overline{\partial}$-lemma \cite{Anderson:2014xha}. The question we will try to address in the following sections is how much of this technology can be taken across to a type IIB setting? In particular we will be interested in what simplifications we observe relative to the heterotic case and what additional complications arise in implementing such an approach to moduli identification. For simplicity, we will confine ourselves in what follows to the closed string sector.

\section{Fluctuation Analysis in Type IIB and Cohomological Interpretation} \label{IIB}

\subsection{Background Solutions} \label{bgsols}

The solutions to the Killing spinor equations for compactifications of type II theories that lead to ${\cal N}=1$ theories with a four-dimensional Minkowski vacuum have been studied in quite some detail (see for example \cite{Grana:2005jc} and references therein). We will consider a spacetime that is a warped product of four-dimensional Minkowski space and an internal manifold admitting an $SU(3)$ structure. Focussing on type IIB string theory, the two 16 component Majorana-Weyl spinors are then decomposed as follows.
\begin{eqnarray} \label{spinoransatz}
\epsilon_1&=&a \, \xi_+ \otimes \eta_+ + \bar{a} \, \xi_- \otimes \eta_- \\ \nonumber
\epsilon_2&=&b \, \xi_+ \otimes \eta_+ + \bar{b} \, \xi_- \otimes \eta_-
\end{eqnarray}
Here $\xi_+$ and $\eta_+$ are positive chirality spinors in four and six dimensions respectively, $\xi_-=\overline{\xi_+}$ and $\eta_-=\overline{\eta_+}$, and $a$ and $b$ are complex functions on the internal six manifold. The vacua that preserve such a set of supersymmetries have been analyzed in great detail using a plethora of different techniques. Here we will only need the results of those analyses in one particular form \cite{Grana:2004bg}.

Given our choice of a warped product of Minkowski space and a compact manifold admitting an $SU(3)$ structure, the metric can be written as
\begin{eqnarray}
ds^2 = e^{2A(y)} \eta_{\mu \nu} dx^{\mu} dx^{\nu} + g_{mn} dy^m dy^n \;,
\end{eqnarray}
where the internal manifold has an associated $J$ and $\Omega$ specifying the $SU(3)$ structure in question.
\begin{eqnarray}
dJ &=& - \frac{3}{2} \textnormal{Im} ( W_1 \overline{ \Omega}) + W_4 \wedge J + W_3 \\ \nonumber
d \Omega &=&  W_1 J \wedge J + W_2 \wedge J + \overline{W}_5 \wedge \Omega
\end{eqnarray}
Here,
\begin{eqnarray}
J \wedge \Omega=0\;,\; W_3 \wedge J =W_3 \wedge \Omega=W_2 \wedge J \wedge J =0
\end{eqnarray}
and $W_1$ is a complex function,  $W_2$ is a complex $(1,1)$ form, $W_3$ is a real form with $(1,2)$ and $(2,1)$ components, $W_4$ is a real one-form, and $W_5$ is a complex $(0,1)$ form.

Fluxes in such a background can be decomposed according to how they transform under the structure group of the internal space. For example, we can write, in a decomposition very similar to that seen for $dJ$ above,
\begin{eqnarray}
H = -\frac{3}{2} \textnormal{Im} (H^{(1)} \overline{\Omega}) + H^{(3)} \wedge J + H^{(6)}\;.
\end{eqnarray}
Here, $H^{(1)}$ is in a singlet representation under the $SU(3)$ structure group, $H^{(3)}$ is in the fundamental representation and $H^{(6)}$ is in the two-index symmetric representation. Similar decompositions can be made for the other form field degrees of freedom that appear in type II theories.

\vspace{0.1cm}

Three special cases, corresponding to restrictions on the possible choices of $a$ and $b$ in (\ref{spinoransatz}), give rise to particularly simple forms for equations for a SUSY preserving vacuum \cite{Grana:2004bg}.

\begin{description}
\item[Case A] If $a=0$ or $b=0$ then we obtain the following conditions. This is the type IIB analogue of the Strominger system solutions of Heterotic string theory.
\begin{eqnarray} \label{eqnsCaseA}
W_1&=&F_3^{(1)} = H^{(1)}_3=W_2 = F_3^{(6)} = \overline{\partial} A = \overline{\partial} a = \overline{\partial} b =0 \\ \nonumber
W_3 &=& \pm * (H_3^{(6)}+H_3^{(\overline{6})}) \\ \nonumber
\overline{W}_5 &=& 2 W_4 = \mp 2 i H_3^{(\overline{3})} = 2 \overline{\partial} \phi
\end{eqnarray}
\item[Case B] if $a= \pm i b$ then we obtain the two following subcases.
\begin{itemize}
\item First, we have the subcase that corresponds to conformally Calabi-Yau solutions
\begin{eqnarray} \label{eqnsCaseBCY}
W_1&=&F_3^{(1)} = H^{(1)}_3 =W_2 =W_3 = \overline{\partial} \phi =0 \\  \label{one}
e^{\phi} F_3^{(6)} &=& \mp * H_3^{(6)} \\ 
e^{\phi} F_5^{(\overline{3})} &=& \frac{2}{3} i \overline{W}_5 = i W_4 = -2 i \overline{\partial} A = -4 i \overline{\partial} \log a
\end{eqnarray}
\item Second, we have the subcase that corresponds to so called `F-theory like' solutions.
\begin{eqnarray} \label{eqnsCaseBF}
W_1&=&F_3^{(1)} = H^{(1)}_3 =W_2 =W_3 = 0 \\  \nonumber
e^{\phi} F_3^{(6)} &=& \mp * H_3^{(6)} \\  \nonumber
e^{\phi} F_1^{(\overline{3})} &=& 2 e^{\phi} F_5^{(\overline{3})} = i \overline{W}_5 = i W_4 = i \overline{\partial} \phi
\end{eqnarray}
\end{itemize}
\item[Cases C] Finally, if $a= \pm b$ then we obtain the following conditions.
\begin{eqnarray} \label{eqnsCaseC}
W_1&=&F_3^{(1)} = H^{(1)}_3=W_2 = H_3^{(6)} =W_4=0 \\ \nonumber
W_3 &=& \pm e^{\phi} * (F_3^{(6)}+F_3^{(\overline{6})}) \\ \nonumber
\pm e^{\phi} F_3^{(\overline{3})} &=& 2 i \overline{W}_5 = -2 i \overline{\partial} A = -4 i \overline{\partial} \log a = -i \overline{\partial} \phi
\end{eqnarray}
\end{description}
Any fluxes or torsion classes that are not mentioned in the above are set to zero in the associated solutions. In all of the cases above there are additional constraints that take the form of primitivity conditions. In each instance, there is one combination of forms which must be $(2,1)$ and primitive. In Case A this is $dJ \pm i H_3$, in Case B $F_3 \mp i e^{-\phi} H_3$ and in Case C $d(e^{-\phi} J) \pm i F_3$.

\vspace{0.1cm}

In addition to the above conditions, in the following sections, we will impose an extra condition upon the compactification manifolds following the analogous constraint that was imposed in \cite{Anderson:2014xha}. We will require that the compact manifold be a  $\partial \overline{\partial}$-manifold. The $\partial \overline{\partial}$-Lemma simply states the following.

\vspace{0.1cm}

{\bf Lemma:} {\it Let $X$ be a compact K\"ahler manifold. For $A$ a d-closed $(p,q)$ form, the following statements are equivalent,}
\begin{eqnarray}
A = \overline{\partial} C \Leftrightarrow A=\partial C' \Leftrightarrow A= d C'' \Leftrightarrow A= \partial \overline{\partial} \tilde{C} \Leftrightarrow A= \partial \hat{C} + \partial \check{C}
\end{eqnarray}
{\it for some $C$, $C'$, $C''$, $\tilde{C}$, $\hat{C}$ and $\check{C}$.}

\vspace{0.1cm}

We then call a manifold, K\"ahler or not, a $\partial \overline{\partial}$-manifold if it satisfies these conditions.

\subsection{Fluctuation analysis and cohomological interpretation} \label{fluct}

The analysis of moduli by fluctuating the equations of {\bf Case A} of the previous subsection is in fact largely identical to the heterotic computation carried out in \cite{Anderson:2014xha}, once one sets the gauge field degrees of freedom to zero. Let us therefore start instead with {\bf Case B}, concentrating initially on the first sub-case, corresponding to conformal Calabi-Yau compactifications.

Combining (\ref{one}) with the fact that $H^{(1)}_3=H^{(3)}_3=F^{(1)}_3=F^{(3)}_3=0$ leads to the following equation.
\begin{eqnarray} \label{two}
e^{\phi} F_3=i(H_{(1,2)}-H_{(2,1)})
\end{eqnarray}
Here $H_{(i,j)}$ corresponds to the components specified in the subscript of the relevant three-form. We will consider the fluctuation of this equation first.

As in Section \ref{het}, we do not wish to write eqn. (\ref{two}) in complex coordinates in order to perform the fluctuation, as the natural complex coordinates will change as we vary the complex structure. We therefore follow the usual procedure of defining projectors,
\begin{eqnarray}  \label{projectors}
\Pi^{(\pm)} = \frac{1}{2} \left(\id \pm i {\cal J} \right)\;,
\end{eqnarray} 
and then rewrite eqn. (\ref{two}) as,
\begin{eqnarray} 
e^\phi F_{lmn}=i(\Pi^{(-)i}_l \Pi^{(-)j}_m \Pi^{(+)k}_n+\Pi^{(-)i}_l \Pi^{(+)j}_m \Pi^{(-)k}_n+\Pi^{(+)i}_l \Pi^{(-)j}_m \Pi^{(-)k}_n \\ \nonumber-\Pi^{(+)i}_l \Pi^{(+)j}_m \Pi^{(-)k}_n-\Pi^{(+)i}_l \Pi^{(-)j}_m \Pi^{(+)k}_n -\Pi^{(-)i}_l \Pi^{(+)j}_m \Pi^{(+)k}_n)H_{ijk}
\;.
\end{eqnarray}
Fluctuating the complex structure ${\cal J}$, the two form potentials $B_2$ and $C_2$, the dilaton and the potential $C_0$, one then obtains the desired result, which we write in terms of the original, unperturbed complex coordinates. From the $(1,2)$ components of equation (\ref{two}) one obtains the following.
\begin{eqnarray} \label{fluc2}
i \delta \tau e^{2\phi} F_{a\overline{b} \overline{c}} +2 \delta {\cal J}_{[\overline{b}}^{\;d} H_{\overline{c}]ad} = (\overline{\partial} \delta \Lambda)_{a \overline{b} \overline{c}}
\end{eqnarray}

The $(0,3)$ component of the fluctuation results in no non-trivial constraint. The $(2,1)$ and $(3,0)$ components of the fluctuation of (\ref{two}) are, of course, simply conjugate to these. The $(0,3)$ component of the fluctuation of the equation $H_3^{(1)}=0$ also results in a non-trivial constraint, which is as follows.
\begin{eqnarray} \label{fluc1}
-\frac{i}{2} \delta {\cal J}_{[\overline{a}}^{\;d} H_{\overline{b}\overline{c}]d}= \overline{\partial}_{[\overline{a}} \left(\delta B_{2\, \overline{b}\overline{c}]}\right) \;.
\end{eqnarray}

In deriving (\ref{fluc2}) and (\ref{fluc1}), we have used the Bianchi Identities for the form fields, which imply in particular that $\delta F_3= d \delta C_2 -\delta C_0 H_3 - C_0 d \delta B_2$, the fact that $\phi$ and $C_0$ are constant in background, the definition of $\tau=C_0 + i e^{-\phi}$ and the $\partial \overline{\partial}$-lemma. The quantity $\delta \Lambda$ is a combination of forms which is implicit due to the use of the $\partial \overline{\partial}$-lemma and whose exact form will not be needed. Note that the fluctuations of the equation $F_3^{(1)}=0$ is built into the above analysis and does not result in any further constraints.

In both equations the left hand side represents a mapping between Dolbeault cohomology groups, as in Section \ref{het}. To show that these maps are well defined, we must show that the left hand sides of equations (\ref{fluc1}) and (\ref{fluc2}) are $\overline{\partial}$ closed, and that shifting $\delta {\cal J}$ by an $\overline{\partial}$ exact piece only changes these combinations by an exact piece. For case B we have from (\ref{eqnsCaseBCY}) that $W_1=W_2=0$, telling us that the manifold is complex and $\delta {\cal J} \in H^1(TX)$.  This, combined with the Bianchi Identity for $H$, is enough to ensure that the left hand side of (\ref{fluc2}) is closed. Similarly, using that $F_1=0$ and $\overline{\partial} \tau=0$, together with the Bianchi Identities for $F$ and $H$, one can easily show that the left hand side of (\ref{fluc2}) is closed. Exactness of the left hand side of these equations under an exact shift in $\delta {\cal J}$ is equally easy to prove and we need only use the Bianchi Identity for $H$ to see this directly. 

Given this analysis, we can see the left hand sides of equations (\ref{fluc1}) and (\ref{fluc2}) as representing maps between Dolbeault cohomology groups. The right hand sides of these equations being exact then tell us that the fluctuations of ${\cal J}$ and $\tau$ which solve the equations of motion are those which are correspond to the kernel of these maps. Thus we have the constraint that allowed moduli of the system must be in the following two kernels.
\begin{eqnarray} \label{ker1}
&&\ker \left( H^1(TX) \stackrel{H_{(1,2)}}{\longrightarrow} H^3(X)\right) \\ \label{ker2}
&&\ker \left( H^0(X) \oplus H^1(TX) \stackrel{F_{(1,2)}, H_{(2,1)}}{\longrightarrow} H^2(TX^{\vee})\right)
\end{eqnarray}
Note here that, since $\overline{\partial} \phi=F_1=0$ we can regard $\tau$ as an element of $H^0(X)$.

One may ask why a constraint such as (\ref{ker1}) does not appear in the Strominger system case \cite{Anderson:2014xha}. After all, we see from (\ref{eqnsCaseA}) that this also has $H_{3\;(3,0)}=0$. The difference occurs because in the Strominger system case $H_{(1,2)}=\frac{i}{2} \overline{\partial}J$. In such a case, one can show \cite{Anderson:2014xha}, that 
\begin{eqnarray}
\delta {\cal J}_{[\overline{a}}^{\; d} H_{\overline{b} \overline{c} ]d} =\frac{i}{2} \delta {\cal J}_{[\overline{a}}^{\; d} (\overline{\partial}J)_{\overline{b} \overline{c} ]d} =- \overline{\partial}_{[\overline{a}} \delta J_{\overline{b}\overline{c}]} \;.
\end{eqnarray}
In such a situation, the analogue of (\ref{fluc1}) is always soluble, being manifestly exact on both sides, and simply becomes an equation that links the fluctuations of certain $(0,2)$ components of fields. In the case at hand, $H$ is related to $e^{\phi}F$ via (\ref{two}), not $\overline{\partial}J$. In such a situation no such simplification can be achieved and an extra constraint is indeed imposed. This distinction will be important in the next section when we match the above map structure to very well known results in Calabi-Yau compactifications of Type IIB string theory.

\vspace{0.1cm}

For the remaining equations in this case, $W_1=W_2=0$ have already been included above, telling us that the perturbations must maintain the complex nature of the compactification manifold. The constraints on $W_3$, $W_4$, $W_5$ and $F_5^{(\overline{3})}$ we expect to correspond to D-term type constraints in the effective theory and, as such, we don't consider these here. We expect this as writing these constraints in terms of fields appearing in the theory, one finds that they all involve contractions with the metric (c.f. (\ref{fandd}) and the surrounding discussion). In addition, we will see further evidence that the F-term constraints are captured by the equations considered above in Section \ref{suprel}.  The equality relating $A$ and $a$ finally, does not affect the physical spectrum directly.

\vspace{0.1cm}

Before moving on to Case C we should briefly mention the second subcase of Case B found in (\ref{eqnsCaseBF}). This case is almost identical in its analysis to the first subcase just considered. This is because the differences between the two cases are largely found in the terms that we expect to be associated to D-terms and thus do not analyse. One exception to this is the constraint on $F_1^{(\overline{3})}$, which may be rewritten as follows.
\begin{eqnarray}
-i (F_{(1,0)}-F_{(0,1)})=d(e^{-\phi})
\end{eqnarray}
Perturbing as before we then find the following constraint.
\begin{eqnarray} \label{newtoF}
\delta {\cal J}_{\overline{a}}^{\;b}F_b =\overline{\partial}_{\overline{a}} \left( \delta \phi e^{-\phi} +i \delta C_0\right)
\end{eqnarray}
Equation (\ref{newtoF}) can be reinterpreted as the following kernel constraint on the complex structure moduli.
\begin{eqnarray} \label{newkerF}
\textnormal{ker} \left( H^1(TX) \to H^1(X)\right)
\end{eqnarray}
Note that in many cases of interest in dimensional reduction one would chose to work on manifolds where $h^1(X)=0$ and in such a case this additional kernel would provide no additional constraint.

\vspace{0.2cm}

The fluctuation of the supersymmetry conditions corresponding to {\bf Case C} follows a similar methodology to the cases discussed above. 
In particular, the equations involving $W_4$, $W_5$ and $F_3^{(\overline{3})}$ all correspond to what we are referring to as `D-term constraints' and so are not considered in this paper. The remaining equations, involving $F_3^{(1)}$, $H_3^{(1)}$, $H_3^{(6)}$ and $F_3^{(6)}$ are of interest to us, however, and we analyze these now. We start with the equation involving $F_3^{(6)}$.
\begin{eqnarray} \label{c1}
W_3 = \pm e^{\phi} * (F_3^{(6)}+ F_3^{(\overline{6})}).
\end{eqnarray}
Using that $W_4=W_1=0$ and that $F_3^{(6)}$ is primitive, we can rewrite (\ref{c1}) as follows.
\begin{eqnarray}
dJ = \pm e^{\phi} i (F_3^{(6)}+ F_3^{(\overline{6})})
\end{eqnarray}
Using the fact that $F_3^{(1)}=0$, we can then obtain
\begin{eqnarray}
dJ_{(2,1)} = \pm i e^{\phi} \left( F_3 - J \wedge F_3^{(3)} \right)_{(2,1)} = \left( \pm i e^{\phi} F_3 +J \wedge \partial \phi \right)_{(2,1)}
\end{eqnarray}
where in the second equality we have used $\pm e^{\phi}F_3^{(3)}=i \partial \phi$ from (\ref{eqnsCaseC}). Performing some elementary algebra we then arrive at the following expression.
\begin{eqnarray} \label{thisbadger}
F_{3\;(2,1)}=\mp i \left(d(e^{-\phi} J)\right)_{(2,1)} \Rightarrow F_3 = \pm i (\overline{\partial}-\partial) (e^{-\phi} J)
\end{eqnarray}
This equation is now in an analogous form to (\ref{two}), and we can analyze its fluctuations in the same manner. From the $(1,2)$ component of the fluctuation one finds the following constraint.
\begin{eqnarray} \label{singleconstraint}
\delta {\cal J}_{[\overline{a}}^{\; d} \left( -i F_{3 \; \overline{b}] dc} \mp \partial_{|d} (e^{-\phi} J)_{c| \overline{b}]}\right) = \mp 2 i \overline{\partial}_{[\overline{a}} \delta ( e^{-\phi} J)_{\overline{b}]c}
\end{eqnarray}
From the $(0,3)$ component we do not obtain another independent constraint. Both sides of the relevant perturbation equation are manifestly exact upon using the equations of motion and we are left with a simple linking of the fluctuations of certain $(0,2)$ components of fields. In a manner analogous to what is seen in the Strominger system case, the equation  $F_3^{(1)}=0$ does not lead to any further constraints once one utilizes (\ref{thisbadger}). It is also easy to see that the equations telling us that $H_3=0$ do not lead to a non-trivial constraint in this case.

In terms of cohomology, our single constraint (\ref{singleconstraint}) can be recast in the following form,
\begin{eqnarray}
\textnormal{ker} \left( H^1(TX) \stackrel{\mp \partial (e^{-\phi} J)}{\longrightarrow} H^2(TX^{\vee}) \right) \;,
\end{eqnarray}
in complete analogy to the examples we have already seen.

\vspace{0.1cm}

It is interesting to note that in all three cases, the map whose source is simply $H^1(TX)$ is defined by the quantity which is primitive in that type of compactification, as described just under (\ref{eqnsCaseC}). This is in direct analogy to what was seen in Section \ref{het} for the case of the Atiyah class. Note also that we would not expect fluctuations to be able to take the system between the different cases listed above. We have not needed to mention the quantization of the background fluxes in the above analysis, but such quantization is indeed in effect in these compact solutions. Since the flux quanta are different in Cases A, B and C and can't be changed under an infinitesimal fluctuation, such cross-talk between these three possibilities should not in general occur.

\subsection{Relationship to Gukov-Vafa-Witten superpotentials} \label{suprel}

It should be noted that the constraints on the allowed field fluctuations satisfying the equations of motion such as (\ref{ker1}) and (\ref{ker2}) are associated with the compactification scale. That is, degrees of freedom not living in these kernels would be expected to have a mass of that magnitude and thus should not be included in a description of the four dimensional effective theory. Nevertheless, there can arise special circumstances (c.f the heterotic case \cite{Anderson:2010mh}) where the mass scale associated to these heavy degrees of freedom is parametrically lower for some reason. In such instances, one can regard these constraints as coming from a Gukov-Vafa-Witten superpotential \cite{Gukov:1999ya} induced mass term. This also clarifies the terminology of  `F-term' and `D-term' constraints that has been employed previously in this paper.

In fact, it is very well known that the susy equations in type II can be derived from the Gukov-Vafa-Witten superpotential (see for example \cite{Koerber:2007xk}), and this fact is of course not changed by writing the potential minimizing degrees of freedom in terms of kernels of maps between Dolbeault cohomology groups. Given this, we will simply content ourselves with showing how a single example, Case B, is reproduced by an analysis of the superpotential and note that the other cases can be obtained in a directly analagous manner.

The relevant superpotential in this case is
\begin{eqnarray}
W \ni \int \left( F_3 -  i e^{-\phi}  H_3 \right) \wedge \Omega \;.
\end{eqnarray}
Note that here we have neglected to include terms proportional to $dJ$. This is because, due to the fact that $W_1=0$ in Case B, $dJ \wedge \Omega=0$ in these examples making this term in the superpotential vanish. This vanishing is preserved under fluctuation. 

Taking the derivative of this superpotential with respect to the four-dimensional axio-dilaton we obtain the following expression.
\begin{eqnarray} \label{dwdtau}
\frac{\partial W}{\partial  \tau} =- \int H_3 \wedge \Omega \;,
\end{eqnarray}
Here we have used the fact that  $\overline{\partial} \phi=F_1=0$ in Case B to isolate the obvious zero mode descending from $\tau$ and have called the resulting four dimensional field by the same name in a slight abuse of notation. Similarly, taking the derivative of the superpotential with respect to the four-dimensional complex structure moduli we obtain the following.
\begin{eqnarray} \label{dwdz}
\frac{\partial W}{\partial z^i} =-\frac{i}{2}\int \left(  \hat{F}_3-\tau H_3 \right) \wedge (v_i \llcorner \Omega)
\end{eqnarray}
Here we have expanded a fluctuation in the complex structure tensor as $\delta {\cal J} = z^i v_i$ where the $v_i$ are a basis of $H^1(TX)$, the field strength $\hat{F}_3$ is the object for which $d\hat{F}_3=0$, and $(v_i \llcorner \Omega)_{\overline{a} bc}\equiv v_{i \, \overline{a}}^{\;\;\;d} \Omega_{dbc}$.

Varying all of the fields in (\ref{dwdtau}) and using all of the same information that was used to derive (\ref{fluc1}) we arrive at the following expression.
\begin{eqnarray} \label{deltadwdtau}
\delta \left(\frac{\partial W}{\partial  \tau} \right)=-\int 3\epsilon^{\overline{a}\overline{b}\overline{c}} \epsilon^{abc}\Omega_{a b c}\left[ \frac{i}{2} \delta {\cal J}_{\overline{a}}^{\; d} H_{d\overline{b}\overline{c}} + \overline{\partial}_{\overline{a}} \left(\delta B_{2 \; \overline{b}\overline{c}} \right)\right]
\end{eqnarray}
Likewise, varying all of the fields in (\ref{dwdz}) we arrive at the following.
\begin{eqnarray} \label{deltadwdz}
\delta \left(\frac{\partial W}{\partial  z^i} \right) =-\frac{i}{2} \int 2 e^{-\phi} \epsilon^{\overline{a}\overline{b}\overline{c}} \epsilon^{abc}\Omega_{a b c} v_{i \; \overline{a}}^a \left[i \delta \tau e^{2\phi} F_{a \overline{b} \overline{c} }+2\delta {\cal J}_{\overline{b}}^{\;d}H_{\overline{c} ad} - (\overline{\partial}\Lambda)_{a\overline{b}\overline{c}}\right]
\end{eqnarray}

We see that asking that a variation of the fields preserves $\frac{\partial W}{\partial \tau}=\frac{\partial W}{\partial z^i}=0$ leads directly to the constraints (\ref{ker1}) and (\ref{ker2}) as expected. This is directly analogous to what was seen in the heterotic case in (\ref{F2}).

\section{Conformal Calabi-Yau Examples With Flux} \label{IIBEGS}

A simple case where the maps in cohomology described in Section \ref{fluct} can be performed explicitly is furnished by the conformal Calabi-Yau compactifications associated to Case B in that Section. There we saw that the subset of the axio-dilaton and complex structure degrees of freedom that are true moduli are given by the following kernels of maps.
\begin{eqnarray} \label{ker1again}
&&\ker \left( H^1(TX) \stackrel{H_{(1,2)}}{\longrightarrow} H^3(X)\right) \\ \label{ker2again}
&&\ker \left( H^0(X) \oplus H^1(TX) \stackrel{F_{(1,2)}, H_{(2,1)}}{\longrightarrow} H^2(TX^{\vee})\right)
\end{eqnarray}
Here, the maps themselves are valued in the sheaf cohomology groups $H^2(TX^{\vee})$ and \newline $(H^2(TX^{\vee}),H^1(\wedge^2 TX^{\vee}))$ respectively.

For generic enough choices of maps in (\ref{ker1again}) and (\ref{ker2again}) one might expect that these kernels will be empty. This is simply the usual statement that one generically expects all of the complex structure and the axio-dilaton to be stabilized by flux \cite{Becker:1996gj,Dasgupta:1999ss,Greene:2000gh,Giddings:2001yu}. In the current context this can be seen by the fact that the dimension of the target spaces in the maps are $1$ and $h^1(TX)$ respectively, and thus one might assume that these maps generically lead to a number of constraints equal to the number of complex structure moduli plus one (for the axio-dilaton). More generally, however, we might wish to know if this generic statement actually holds true for a particular flux, and if not which moduli are stabilized and which are not fixed. It is to this question that we turn in specific examples in this section.

We can explicitly describe the various spaces involved in the computation, as detailed above, using standard techniques from computational algebraic geometry.  In this paper we will illustrate this with examples based on complete intersections in products of projective spaces, or CICYs \cite{Yau:1986gu,Hubsch:1986ny,Green:1986ck,Candelas:1987kf,Candelas:1987du}\footnote{The type of computations being considered here could easily be extended to the case of generalized CICYs \cite{Anderson:2015iia}. See \cite{Berglund:2016yqo,Berglund:2016nvh,Garbagnati:2017rtb} for related work. Many of the computations in the following were carried out using the ``CICY Package" \cite{cicy}}. Similar techniques could easily be applied in any case where one has enough control over the relevant cohomology groups. A CICY is described by a configuration matrix of the following form.
\begin{eqnarray} \label{cicy}
M_X=\left[ \begin{array}{c|ccc} n_1 & q_1^1 & \ldots & q_K^1 \\ \vdots&\vdots & \ddots&\vdots \\n_m & q_1^m & \ldots & q_K^m \end{array}\right]
\end{eqnarray}
Here, the first column in $M_X$ describes an ambient product of $m$ projective spaces, $\mathbb{P}^{n_1} \times \ldots \times \mathbb{P}^{n_m}$. The remaining columns each describe one of $K$ defining equations which specify the Calabi-Yau manifold within the ambient space. The integers $q$ specify the multi degree of each defining relation in terms of the homogeneous coordinates of the ambient project space factors.

In order to obtain a description of the cohomologies appearing in (\ref{ker1again}) and (\ref{ker2again}) on a CICY of the form (\ref{cicy}), we will make use of the following exact sequences.
\begin{itemize}
\item The adjunction sequence,
\begin{eqnarray}
0 \to TX \to TA|_X \to N_X \to 0 \;.
\end{eqnarray}
\item The Euler sequence for the tangent bundle to the ambient product of projective spaces, restricted to the Calabi-Yau.
\begin{eqnarray}
0 \to {\cal O}_X^m \to {\cal O}_X(1,0,\ldots,0)^{\oplus n_1} \ldots \oplus {\cal O}_X(0,0,\ldots,1)^{\oplus n_k} \to TA|_X \to 0
\end{eqnarray}
\item The Koszul sequence relating sheaves over the ambient space and sheaves over the Calabi-Yau threefold.
\begin{eqnarray}
0 \to {\cal V} \otimes \wedge^k N^{\vee} \to {\cal V} \otimes \wedge^{k-1} N^{\vee} \to \ldots \to {\cal V} \wedge N^{\vee} \to {\cal V} \to {\cal V}|_X \to 0
\end{eqnarray}
\item The exterior power sequence which is defined as follows. Given a short exact sequence:
\begin{eqnarray}
0 \to A\to B\to C \to0
\end{eqnarray}
the exterior power sequence is given by
\begin{eqnarray} \label{extpower}
0 \to S^k A \to S^{k-1} A \otimes B \to S^{k-2} A \otimes \wedge^2 B \to \ldots \to \wedge^k B \to \wedge^k C \to 0
\end{eqnarray}
for any $k$. A similar sequence exists with the symmetric and antisymmetric products interchanged. 
\item The dual sequences of all those listed above.
\end{itemize}
Splitting these sequences up into short exact pieces using kernels and cokernels, we can then take the associated long exact sequences in cohomology. Sequence chasing can then be used to relate the cohomologies of interest to simply ambient space line bundle cohomologies. These in turn can then be described by use of the theorem due to Bott, Borel and Weil \cite{Hubsch:1992nu}. Finally, in the examples we give, we will consider smooth quotients of CICYs rather than CICYs themselves in order to facilitate computation. We will thus be interested in the invariant parts of these cohomology groups under the group action induced on them from the quotiented symmetry.

Below, we will illustrate all of this with two examples. For simplicity, we begin with an example which is a freely acting quotient of the famous quintic Calabi-Yau threefold.

\subsection{A simple quintic example}

We will begin with a simple example defined as a quotient of the quintic Calabi-Yau threefold, described by the following configuration matrix,
\begin{eqnarray} \label{conf}
M_X = \left[ \begin{array}{c|c} \mathbb{P}^4 & 5 \end{array}\right]
\end{eqnarray} 
by a freely acting $ \mathbb{Z}_5 \times \mathbb{Z}_5$ symmetry. The configuration matrix (\ref{conf}) indicates that $X$ is defined as the zero locus of a degree 5 polynomial inside $\mathbb{P}^4$. We will denote the homogeneous coordinates on $\mathbb{P}^4$ as $x_i$ where $i=0,\ldots 4$. The freely acting $ \mathbb{Z}_5 \times \mathbb{Z}_5$ symmetry by which we will quotient $X$ has generators given by 
\begin{eqnarray} 
g_1: x_i \to \omega^{ i} x_i \\ \nonumber
g_2: x_i \to x_{i+1}
\end{eqnarray}
where $\omega$ is a fifth root of unity and we define $x_{5}=x_0$. The quotient manifold $X/ \mathbb{Z}_5 \times \mathbb{Z}_5$ is a smooth Calabi-Yau threefold, for sufficiently generic choices of complex structure, and has $h^{1,1}=1$ and $h^{2,1}=5$ \cite{Candelas:2008wb,Candelas:2010ve,Candelas:2015amz,Candelas:2016fdy,Constantin:2016xlj}.

Using the sequences mentioned at the start of this section, one can compute that the complex structure moduli are encoded by the following description of the first tangent bundle valued cohomology group.
\begin{eqnarray} \label{h1tx1}
H^1(TX) = \frac{\textnormal{Coker} \left[ \mathbb{C} \to [5] \right]}{\textnormal{Coker} \left[ \mathbb{C} \to [1]^{\oplus 5} \right]}
\end{eqnarray}
Here the map used in defining the quotient is given by $dP$ (the derivative of the defining relations), that in the numerator is given by $P$ (the defining relations themselves) and that in the denominator is given by the homogeneous coordinates of the ambient space. The symbols $[n]$ where $n$ is an integer denote the spaces of polynomials of degree $n$. 
More precisely, they are those such polynomials that are invariant under the $\mathbb{Z}_5 \times \mathbb{Z}_5$ action.

Similarly we have that
\begin{eqnarray} \label{h2txv1}
H^2(TX^{\vee}) = \textnormal{Ker} \left[ \textnormal{Ker}[ [-5] \to \mathbb{C}] \stackrel{dP}{\longrightarrow} \textnormal{Ker} [ [-1]^{\oplus 5} \to \mathbb{C} ] \right]
\end{eqnarray}
where the map in the first kernel is given by the defining relation and that in the second is given by the homogeneous coordinates. The symbols $[n]$ where $n$ is a negative integer here denote spaces of rational functions of a given degree. More precisely, $[-|n|]$ denotes the space of rational functions constructed as a sum of terms, each of which is a rational monomial of the given degree. As in the $[|n|]$ case, only those functions that are invariant under the group action are included.

Finally, we will require the following description of this tangent bundle valued cohomology.
\begin{eqnarray} \label{mapspace}
H^1 (\wedge^2 TX^{\vee}) = \textnormal{Ker} \left[ \textnormal{Ker} \left\{ [-10]\to [-5] \right\} \to \textnormal{Ker} \left[ \textnormal{Ker} \left\{ [-6] \to [-1] \right\}^{\oplus 5} \to \textnormal{Ker} \left\{ [-5] \to \mathbb{C} \right\} \right] \right]
\end{eqnarray}
Once more the maps in this expression are described by $P$, $dP$ and the homogeneous coordinates, with which map is to be used being determined by which has the appropriate by degree.

\vspace{0.1cm}

With these descriptions of the relevant cohomologies in hand, let us proceed to compute the first kernel, given in (\ref{ker1again}). First, for simplicity in this initial example, we will choose our defining relation to be the Fermat quintic.
\begin{eqnarray} \label{def}
P= x_0^5+x_1^5+x_2^5+x_3^5+x_4^5
\end{eqnarray} 
We take a general element of $H^1(TX)$, as described by (\ref{h1tx1}).
\begin{eqnarray}  \label{gensource}
c_1 x_0 x_1 x_2 x_3 x_4+c_2 \left(x_3 x_4^2 x_2^2+x_0^2 x_1 x_2^2+x_1^2 x_3^2 x_2+x_0 x_1^2 x_4^2+x_0^2 x_3^2 x_4\right) \\ \nonumber +c_3 \left(x_0^2 x_3 x_1^2+x_2^2 x_4 x_1^2+x_3^2 x_4^2 x_1+x_0 x_2^2 x_3^2+x_0^2 x_2 x_4^2\right)+c_4 \left(x_2 x_3 x_0^3+x_1 x_3^3 x_0+x_2^3 x_4 x_0 \right. \\ \nonumber  \left.+x_1 x_2 x_4^3+x_1^3 x_3 x_4\right)+c_5 \left(x_1 x_4 x_0^3+x_3 x_4^3 x_0+x_1^3 x_2 x_0+x_1 x_2^3 x_3+x_2 x_3^3 x_4\right) 
\end{eqnarray}
Note that this description of $H^1(TX)$ is in terms of degree five polynomials. As such this formulation of the complex structure is extremely easy to interpret. The elements of this space which lie in both of the kernels (\ref{ker1again}) and (\ref{ker2again}) are the fluctuations of the complex structure moduli allowed by the equations of motion. Small multiples of these polynomials can then be added to the initial defining relation (\ref{def}) to see which family of Calabi-Yau hypersurfaces is left unstabilized by the given choice of fluxes.

To perform the mapping in (\ref{ker1again}) we will need a choice of flux $H_{(1,2)}$. This should be an element of $H^2(TX^{\vee})$ and thus we describe it as in (\ref{h2txv1}). In fact, we should be cautious as the flux we choose should be primitive according to the supergravity equations of motion ($F_3 \mp i e^{-\phi}H_3$ is primitive as mentioned in Section \ref{bgsols} and the relevant components of $F_3$ and $H_3$ are proportional as seen in (\ref{two})). Fortunately a big simplification occurs here with respect to the heterotic case. In the heterotic examples of Section \ref{hetegs}, it is not guaranteed that for a choice of map cohomology class, a poly-stable holomorphic vector bundle exists whose field strength gives rise to that map. For fluxes in type IIB string theory, however, the situation is quite different.

Consider a $(2,1)$ field strength in any given cohomology class. The question we wish to know the answer to is, is there a field strength in the same cohomology class which is primitive? That is, if we have ${\cal H}$ such that $[{\cal H}] \in H^{2,1}(X)$ does there exist ${\cal H}'$ satisfying $[{\cal H}'] =[{\cal H}]$ such that ${\cal H}' \wedge J =0$?

For the case at hand, that where $X$ is a Calabi-Yau threefold, ${\cal H} \wedge J$ is an element of $H^{3,2}(X)$. Since $h^{3,2}(X)=0$ for such a manifold we know that ${\cal H} \wedge J = \overline{\partial} \Lambda$ for some four-form $\Lambda$. In fact, we know a little more than this, thanks to some very well know results.

The Hard Lefschetz theorem states that the map
\begin{eqnarray}
L^k:H^{d-k}(X) \to H^{d+k}(X)
\end{eqnarray}
is an isomorphism. Here $d$ is the complex dimension of $X$ and $L$ is the map on cohomology induced by the operation of performing a wedge product with the K\"ahler form. Taking the case where $k=2$ we see that $L^2: H^1(X) \to H^{5}(X)$ is an isomorphism. That is, any element of $H^5(X)$ can be written as $J \wedge J \wedge \alpha$ for some $\alpha$. This implies that the same is true for $H^{3,2}(X)$ and thus, in the notation of the proceeding paragraph, $\overline{\partial} \Lambda= J \wedge J \wedge \overline{\partial} \gamma$ for some function $\gamma$ (using the fact that $h^1(X)=0$).

Using this information we can now easily see the desired result. By definition, ${\cal H}' = {\cal H} + \overline{\partial} \beta$ for some two-form $\beta$. Then $J \wedge {\cal H}' = J \wedge {\cal H} + J \wedge \overline{\partial}\beta = J \wedge J \wedge \overline{\partial} \gamma + J \wedge \overline{\partial} \beta$ and we see that an appropriate choice of $\beta$ (namely $\beta= - J \wedge \gamma$) renders ${\cal H}'$ primitive as desired.

Thus, which ever class in $H^2(TX^{\vee})$ we choose, a suitable choice of primitive flux will exist within that class. An exactly analogous argument can be made for $H^1(\wedge^2 TX^{\vee})$. The kernel computations we are performing here only depend upon the class of the map elements being used, and as such we do not need to know the exact form of the primitive representative to proceed.

\vspace{0.1cm}

In the case at hand, the most general possible map appearing in (\ref{ker1again}), as described by (\ref{h2txv1}) and depending upon $h^2( TX^{\vee})=5$ parameters $m_{\alpha}$, is given explicitly as follows.
\begin{eqnarray} \label{h2txv}
\frac{m_1}{x_0 x_1 x_2 x_3 x_4}+m_2 \left(\frac{1}{x_1^2 x_2 x_3^2}+\frac{1}{x_0^2 x_3^2 x_4}+\frac{1}{x_0 x_1^2 x_4^2}+\frac{1}{x_2^2 x_3 x_4^2}+\frac{1}{x_0^2 x_1 x_2^2}\right) \\ \nonumber+m_3 \left(\frac{1}{x_0 x_2^2 x_3^2}+\frac{1}{x_1^2 x_2^2 x_4}+\frac{1}{x_0^2 x_2 x_4^2}+\frac{1}{x_1 x_3^2 x_4^2}+\frac{1}{x_0^2 x_1^2 x_3}\right) \\\nonumber+m_4 \left(\frac{1}{x_0 x_1 x_3^3}+\frac{1}{x_0 x_2^3 x_4}+\frac{1}{x_1^3 x_3 x_4}+\frac{1}{x_1 x_2 x_4^3}+\frac{1}{x_0^3 x_2 x_3}\right)\\ \nonumber+m_5 \left(\frac{1}{x_1 x_2^3 x_3}+\frac{1}{x_0^3 x_1 x_4}+\frac{1}{x_2 x_3^3 x_4}+\frac{1}{x_0 x_3 x_4^3}+\frac{1}{x_0 x_1^3 x_2}\right)
\end{eqnarray}

We multiply such a map by the general source element given in (\ref{gensource}) and trim the result to only include constants - the relevant description of the target space in (\ref{ker1again}), $H^3(X)$. We find that, for a fluctuation of the form (\ref{gensource}) to appear in the kernel (\ref{ker1again}) the following constraint on the coefficients $c_i$ must hold.
\begin{eqnarray} \label{constr1}
c_1 m_1+5 c_2 m_2+5 c_3 m_3+5 c_4 m_4+5 c_5 m_5=0
\end{eqnarray}

So for example, if we choose the map corresponding to $m_1=5,m_2=3,m_3=4,m_4=10$ and $m_5=6$, then the most general fluctuation of the defining relation of the quotiented quintic which is allowed by the first constraint (\ref{ker1again}) is as follows.
\begin{eqnarray}
c_2 \left(x_3 x_4^2 x_2^2+x_0^2 x_1 x_2^2+x_1^2 x_3^2 x_2-3 x_0 x_1 x_3 x_4 x_2+x_0 x_1^2 x_4^2+x_0^2 x_3^2 x_4\right) \\ \nonumber+c_3 \left(x_0^2 x_3 x_1^2+x_2^2 x_4 x_1^2+x_3^2 x_4^2 x_1-4 x_0 x_2 x_3 x_4 x_1+x_0 x_2^2 x_3^2+x_0^2 x_2 x_4^2\right)\\\nonumber+c_4 \left(x_2 x_3 x_0^3+x_1 x_3^3 x_0+x_2^3 x_4 x_0-10 x_1 x_2 x_3 x_4 x_0+x_1 x_2 x_4^3+x_1^3 x_3 x_4\right)\\\nonumber+c_5 \left(x_1 x_4 x_0^3+x_3 x_4^3 x_0+x_1^3 x_2 x_0-6 x_1 x_2 x_3 x_4 x_0+x_1 x_2^3 x_3+x_2 x_3^3 x_4\right)
\end{eqnarray}
We see that we get one constraint on the general five parameter possible complex structure fluctuation as should be the case.

\vspace{0.1cm}

As we have seen, the first map is easily implemented, and the constraint on moduli it corresponds to can be mapped out explicitly. We now move on to consider the second kernel condition (\ref{ker2again}). Here we will see a complication in comparison to the heterotic case. 

In the case of the second kernel condition the source space is the direct sum of (\ref{gensource}), the complex structure fluctuations (which must also be constrained by the first condition), and the constants (which is the relevant description of $H^0(X)$). The target space is described by an expression of the form (\ref{h2txv}). Finally, the map is described by an element of $H^2(TX^{\vee})$ as in (\ref{h2txv}) (which maps the $H^0(X)$ piece of the source to the target) together with an element of $H^1(\wedge^2 TX^{\vee})$ (which maps the $H^1(TX)$ piece of the source to the target). The relevant description of this last cohomology group, depending upon $h^1(\wedge^2 TX^{\vee})=5$ parameters $n_{\alpha}$, is given explicitly as follows.
\begin{eqnarray} \label{map2space}
n_1 \left(\frac{1}{x_0^3 x_2^3 x_3^2 x_4^2}+\frac{1}{x_0^2 x_1^3 x_3^3 x_4^2}+\frac{1}{x_0^2 x_1^2 x_2^3 x_4^3}+\frac{1}{x_1^3 x_2^2 x_3^2 x_4^3}+\frac{1}{x_0^3 x_1^2 x_2^2 x_3^3}\right) \\ \nonumber+n_2 \left(\frac{1}{x_0^3 x_1^3 x_2^2 x_4^2}+\frac{1}{x_1^2 x_2^3 x_3^3 x_4^2}+\frac{1}{x_0^3 x_1^2 x_3^2 x_4^3}+\frac{1}{x_0^2 x_2^2 x_3^3 x_4^3}+\frac{1}{x_0^2 x_1^3 x_2^3 x_3^2}\right)  \\ \nonumber +n_3 \left(\frac{1}{x_0^2 x_1 x_2^3 x_3^3 x_4}+\frac{1}{x_0 x_1^3 x_2^3 x_3 x_4^2}+\frac{1}{x_0^3 x_1 x_2^2 x_3 x_4^3}+\frac{1}{x_0 x_1^2 x_2 x_3^3 x_4^3}+\frac{1}{x_0^3 x_1^3 x_2 x_3^2 x_4}\right) \\ \nonumber
+n_4 \left(\frac{1}{x_0 x_1^3 x_2^2 x_3^3 x_4}+\frac{1}{x_0^3 x_1 x_2 x_3^3 x_4^2}+\frac{1}{x_0^2 x_1^3 x_2 x_3 x_4^3}+\frac{1}{x_0 x_1 x_2^3 x_3^2 x_4^3}+\frac{1}{x_0^3 x_1^2 x_2^3 x_3 x_4}\right)\\ \nonumber +\frac{n_5}{x_0^2 x_1^2 x_2^2 x_3^2 x_4^2} 
\end{eqnarray}
The complication here arises in that the two components of this map can not be chosen independently from the map, already specified in (\ref{ker1again}). The component living in $H^2(TX^{\vee})$ should be proportional to the map already chosen and so this is relatively easy to determine. The component living in $H^1(\wedge^2 TX^{\vee})$ is more problematic. In terms of differential forms, this map component should be the complex conjugate of the one appearing in (\ref{ker1again}). The problem is that these cohomologies are being described here in algebro-geometric terms and this process of complex conjugation is not transparent in such a formulation. Thus it is rather difficult to know which component of $H^1(\wedge^2 TX^{\vee})$ should be selected. Such a complication does not arise in the heterotic Atiyah class setting where there is a single map composed of a single component and no complex conjugation is required. 

Note that this obstruction can be overcome in cases where the metric on complex structure moduli space \cite{Candelas:1990pi} is known for the Calabi-Yau in question in an appropriate form (see for example \cite{Candelas:1990rm}). In such an instance one can combine this knowledge with the natural pairing
\begin{eqnarray} \label{pairing}
H^1(\wedge^2 TX^{\vee}) \times H^2(TX^{\vee}) \to \mathbb{C}
\end{eqnarray}
in order to isolate the correct conjugate pairing. The point is that this pairing and the metric are essentially the same quantity up to an overall scale, and (\ref{pairing}) can be computed explicitly for the polynomial descriptions of the cohomologies being utilized in this section. If the complex structure moduli space metric is given in bases for the barred and unbarred indices that are known to be conjugate, then the problem becomes soluble by performing a basis change on the cohomological spaces to match the pairing (\ref{pairing}) with that metric. Such an involved computation is beyond the scope of this paper, and indeed it is dissatisfying that one needs to compute a K\"ahler potential in order to learn about flat directions of a superpotential in this approach. Nevertheless, in a case where one wished to have full control of the low energy theory, the K\"ahler potential would be required anyway and the above obstruction would be naturally overcome.

In short, one can easily determine the constraint from either (\ref{ker1again}) or (\ref{ker2again}) where the dilaton is taken to be fixed, but a complete analysis combining both constraints would require this more subtle information. Let us give an example of the type of constraint on pure complex structure fluctuations that can arise from (\ref{ker2again}) in order to illustrate the more complex moduli stabilization results that can arise in this setting. 

Let us choose as an example of the class of the $(2,1)$ field strength in (\ref{ker2again}) the element of (\ref{map2space}) where $n_3=n_4=n_5=0$ and $n_1=n_2=1$. Then, performing the map from $H^1(TX)$ to $H^2(TX^{\vee})$ following a methodology analogous to that described above, we find the following for the allowed fluctuations of the defining relation (that is the complex structure).
\begin{eqnarray} 
P_{\textnormal{fluctuation}}=x_0^5+x_1^5+x_2^5+x_3^5+x_4^5+\delta x_0 x_1 x_2 x_3 x_4
\end{eqnarray}
Here $\delta$ is the fluctuation parameter. We see that this map does not give a generic result for the number of unconstrained moduli. The target space $H^2(TX^{\vee})$ is of the same dimension as $H^1(TX)$ and thus we might naively expect all of the complex structure to be stabilized. With the above direct computation for this choice of map, however, we can see that this is not the case.


\subsection{A more complex example}

In this subsection we will present a slightly more complex example of the kernel computations we have been discussing. This will illustrate both that the approach being presented is not restricted to quotients of the quintic, and some more of the features of these analyses.

We will consider a quotient of the following CICY, number $7669$ in the canonical list \cite{Yau:1986gu,Hubsch:1986ny,Green:1986ck,Candelas:1987kf,Candelas:1987du},
\begin{eqnarray}
M_X = \left[ \begin{array}{c|ccc} \mathbb{P}^2 & 1&1 &1 \\\mathbb{P}^2 & 1&1 &1 \\\mathbb{P}^2 & 1&1 &1   \end{array}\right] \;.
\end{eqnarray}
We will quotient by a freely acting $\mathbb{Z}_3$ symmetry \cite{Candelas:2008wb,Braun:2010vc,Candelas:2010ve}. Defining $x_{a,i}$ to be the homogeneous coordinates on the $a$'th $\mathbb{P}^1$ factor of the ambient space, the group action is defined as follows.
\begin{eqnarray} \label{z3action}
g: x_{a,i} \to x_{a+1,i}
\end{eqnarray}
Here we define $x_{4,i}=x_{1,i}$. The quotient manifold $X/\mathbb{Z}_3$ is a smooth Calabi-Yau threefold, for sufficiently generic choices of complex structure, and has $h^{1,1}= 1$ and $h^{2,1}=16$ \cite{Candelas:2008wb,Candelas:2010ve,Candelas:2015amz,Candelas:2016fdy,Constantin:2016xlj}.

Using the sequences mentioned at the start of this section, one can compute the complex structure moduli are encoded by the following description of the first tangent bundle valued cohomology group.
\begin{eqnarray}  \label{newnew1}
H^1(TX) = \frac{\textnormal{Coker} \left[ \mathbb{C}^{\oplus 9} \to [1,1,1]^{\oplus 3}\right]}{\textnormal{Coker} \left[ \mathbb{C}^{\oplus 3} \to [1,0,0]^{\oplus 3} \oplus [0,1,0]^{\oplus 3} \oplus [0,0,1]^{\oplus 3} \right]}
\end{eqnarray}
Here the map defining the quotient is given by the derivative of the defining equations $dP$, the map in the numerator is determined by the defining relations $P$ themselves, and the map in the denominator is determined by the homogeneous coordinates of the ambient space projective space factors.

Another cohomology that we will need is $H^2(TX^{\vee})$, for which we find the following description.
\begin{eqnarray} \label{newnew2}
H^2(TX^{\vee}) = \textnormal{Ker} \left[ \textnormal{Ker} \left([-1,-1,-1]\stackrel{P}{\longrightarrow} [0,0,0]^{\oplus 3}]\right)^{\oplus 3}\stackrel{dP}{\longrightarrow} \textnormal{Ker}\left[ [-1,0,0]^{\oplus 3} \oplus  \right. \right.\\\nonumber \left. \left.  [0,-1,0]^{\oplus 3} \oplus [0,0,-1]^{\oplus 3} \stackrel{x}{\longrightarrow} \mathbb{C}^{\oplus 3}\right]\right]
\end{eqnarray}
Since the analysis is becoming repetitive we have simply denoted the quantities relevant to this expression over the maps that they define.

The final cohomology, whose description we require to define the map in (\ref{ker2again}) has a more complex description, given as follows.
\begin{eqnarray} \label{bigone}
H^1(\wedge^2 TX^{\vee}) = \textnormal{Ker} \left[ \textnormal{Ker} \left[ ([-2,-2,-2]\to [-1,-1,-1]^{\oplus 3})^{\oplus 6} \to \right.\right.\\\nonumber \left.\left. \textnormal{Ker}\left({\cal K}_1 \to \textnormal{Ker}\left[[-1,-1,-1]\to [0,0,0]^{\oplus 3}\right]^{\oplus 9}\right)\right] \to \mathbb{C}^{\oplus 6} \right]
\end{eqnarray}
In the above, ${\cal K}_1$ is defined by the following set of kernels.
\begin{eqnarray}
{\cal K}_1=(\textnormal{ker} \left[[-2,-1,-1]\to[-1,0,0]^{\oplus 3}\right])^{\oplus 9} \oplus(\textnormal{ker} \left[[-1,-2,-1]\to[0,-1,0]^{\oplus 3}\right])^{\oplus 9}\\\nonumber \oplus(\textnormal{ker} \left[[-1,-1,-2]\to[0,0,-1]^{\oplus 3}\right])^{\oplus 9} 
\end{eqnarray}
While most of the maps in this expression can be obviously constructed from $P$ and $dP$, the final map, to $\mathbb{C}^6$ is more subtle in nature and deserves a little further discussion. This final map in (\ref{bigone}) is, of course, crucial in obtaining a correct description of $H^1(\wedge^2 TX^{\vee})$. In particular taking the kernel of this map removes precisely one degree of freedom to correctly leave a $16$ dimensional space. Unlike the other maps that appear in these expressions, however, it may not be immediately obvious how it is to be constructed. It maps sets of polynomials of degree $[-2,-2,-2]$ to constants but is not built out of products of two defining relations as one might naively expect from the degrees in (\ref{bigone}) (such a map is in fact the trivial map between the two relevant spaces here). 

The space $\textnormal{Ker} \left[ ([-2,-2,-2]\to [-1,-1,-1]^{\oplus 3})^{\oplus 6}\right]$ arises from $H^3(X,S^2 {\cal N}^{\vee})$ in the sequence chasing, whereas the $\mathbb{C}^{\oplus 6}$ target arises from $H^0(X,S^2 ( {\cal O}_X^{\oplus 3}))$. The maps separating these two quantities, from which we must construct the composite map that appears in (\ref{bigone}), are built from the data $dP$ and $x$ (the derivatives of the defining relations and the homogeneous coordinates of the ambient space). Examining the structure of the sequence (\ref{extpower}) applied to this example, we expect the quantity $dP$ to appear quadratically in the resulting composite map. Note that we can not simply combine $dP$ and $x$ in the obvious manner to obtain polynomials of degree $[1,1,1]$ that can then be combined quadratically to define the map. Such quantities are quadratic in defining relations $P$ and thus, given the structure of the kernel $\textnormal{Ker} \left[ ([-2,-2,-2]\to [-1,-1,-1]^{\oplus 3})^{\oplus 6}\right]$, these act as the zero map on this space.

We require, then, a map defined, loosely speaking, from $H^3(X,S^2 {\cal N}^{\vee})$ to $H^0(X,S^2 ( {\cal O}_X^{\oplus 3}))$ using simply homogeneous coordinates of the ambient space and $dP$. To describe this map we will need several pieces. We define a tensor of numbers, $dP_k^{(\alpha) ij}$, as the coefficients of the derivative of the defining relations.
\begin{eqnarray} \label{dPcoeff}
dP_k^{(\alpha)} = dP_k^{(\alpha) ij} x_i x_j
\end{eqnarray}
Here $i,j,k$ are now composite indices that run over all of the homogeneous coordinates of the ambient space and $\alpha$ runs over the defining relations themselves. We also define the antisymmetric tensors $\epsilon^{(A)}_{ijk}$. Here, the index $A$ runs over the projective space factors of the ambient space and the tensor is zero unless all indices $i,j,k$ are taken from the associated projective space and is the Levi-Civita symbol in that case.

With these definitions in place, we can define the map using the following objects.
\begin{eqnarray} \label{image}
\sum_{k_1,k_2} \psi^{k_1 \ldots k_6} \; \epsilon^{(A)}_{i_1i_2 k_3} \epsilon^{(B)}_{j_1j_2k_4} dP_{k_5}^{(\alpha)i_1 j_1} dP_{k_6}^{(\beta)i_2 j_2}
\end{eqnarray}
The quantity given in (\ref{image}) constitutes a piece of the image of the map. The $\psi$ are the coefficients of the degree $[-2,-2,-2]$ polynomials of the source written as a tensor, similarly to what was seen for $dP$ in (\ref{dPcoeff}). Each of the ${\cal O}_X$ in $H^0(X,S^2 ( {\cal O}_X^{\oplus 3}))$ corresponds to one of the ambient space factors. Depending upon which piece of this cohomology we are describing the map to, we pick $A$ and $B$ accordingly. Similarly the source space of $H^3(X,S^2 {\cal N}^{\vee})$ breaks up into pieces labeled by two defining relation counting indices. Depending upon which piece of this cohomology we are describing the map from, we pick $\alpha$ and $\beta$ accordingly. Combining all of these pieces together, a full mapping can be obtained.

Using this map we can show that the image obtained is correctly equivariant. In addition, the further sequence chasing reveals that the cokernel of this map should give a description of $H^2(\wedge^2 TX^{\vee})=H^{2,2}$, that is, the cohomology associated to the single K\"ahler modulus of the quotiented manifold. The cohomology describing the K\"ahler modulus, obtained from the map described above, is then built from a sum of identical pieces, one for each projective space factor, as the nature of the group action (\ref{z3action}) would suggest. Finally, taking the kernel does indeed remove one degree of freedom giving rise to a 16 parameter description of the cohomology $H^1(\wedge^2 TX^{\vee})$ which is invariant under the $\mathbb{Z}_3$ group action induced from (\ref{z3action}).	

\vspace{0.1cm}

Given the descriptions given in equations (\ref{newnew1}),(\ref{newnew2}) and (\ref{bigone}) of the necessary cohomologies one can now analyze the allowed moduli fluctuations, as described by equations (\ref{ker1again}) and (\ref{ker2again}) in exactly the same way as was discussed in the previous example. For example, one can pick the following initial defining relations, which can be easily shown to define a smooth Calabi-Yau threefold by employing standard methods \cite{Candelas:1987kf}.
\begin{eqnarray} \nonumber \label{rancschoice}
P_1&=&x_{1,0} x_{2,0} x_{3,0}+x_{1,1} x_{2,0} x_{3,0}+x_{1,2} x_{2,0} x_{3,0}+x_{1,0} x_{2,1} x_{3,0}+x_{1,1} x_{2,1} x_{3,0}+x_{1,0} x_{2,2} x_{3,0} \\ \nonumber &&+x_{1,2} x_{2,2} x_{3,0}+x_{1,0} x_{2,0} x_{3,1}+x_{1,1} x_{2,0} x_{3,1}+x_{1,0} x_{2,1} x_{3,1}+x_{1,1} x_{2,1} x_{3,1}+x_{1,2} x_{2,2} x_{3,1} \\ \nonumber &&+x_{1,0} x_{2,0} x_{3,2}+x_{1,2} x_{2,0} x_{3,2}+x_{1,2} x_{2,1} x_{3,2}+x_{1,0} x_{2,2} x_{3,2}+x_{1,1} x_{2,2} x_{3,2} \\ \nonumber
P_2&=&x_{1,1} x_{2,2} x_{3,0}+x_{1,2} x_{2,0} x_{3,1}+x_{1,2} x_{2,1} x_{3,1}+x_{1,1} x_{2,2} x_{3,1}+x_{1,0} x_{2,1} x_{3,2}+x_{1,1} x_{2,1} x_{3,2} \\  &&+x_{1,2} x_{2,2} x_{3,2} \\ \nonumber
P_3&=&x_{1,2} x_{2,0} x_{3,0}+x_{1,1} x_{2,1} x_{3,0}+x_{1,0} x_{2,2} x_{3,0}+x_{1,1} x_{2,2} x_{3,0}+x_{1,2} x_{2,2} x_{3,0}+x_{1,1} x_{2,0} x_{3,1} \\ \nonumber &&+x_{1,2} x_{2,0} x_{3,1}+x_{1,0} x_{2,1} x_{3,1}+x_{1,2} x_{2,1} x_{3,1}+x_{1,1} x_{2,2} x_{3,1}+x_{1,0} x_{2,0} x_{3,2}+x_{1,2} x_{2,0} x_{3,2} \\ \nonumber &&+ x_{1,0} x_{2,1} x_{3,2}+x_{1,1} x_{2,1} x_{3,2}+x_{1,0} x_{2,2} x_{3,2}
\end{eqnarray}

Computing the form of a general element of the description of $H^1(TX)$ given in (\ref{newnew1}), we then find the following parametrization of the infinitesimal complex structure fluctuations.
\begin{eqnarray} \label{csfluctbigone}
\delta P_1 &=& s_{16}\,x_{1,0} x_{2,0} x_{3,0}+s_{15} \left(x_{1,1} x_{2,0} x_{3,0}+x_{1,0} x_{2,1} x_{3,0}+x_{1,0} x_{2,0} x_{3,1}\right) \\\nonumber &&+s_{13} \left(x_{1,1} x_{2,1} x_{3,0}+x_{1,1} x_{2,0} x_{3,1}+x_{1,0} x_{2,1} x_{3,1}\right)+s_{14} \left(x_{1,2} x_{2,0} x_{3,0}+x_{1,0} x_{2,2} x_{3,0}+x_{1,0} x_{2,0} x_{3,2}\right)\\\nonumber &&+s_{12} \left(x_{1,2} x_{2,1} x_{3,0}+x_{1,0} x_{2,2} x_{3,1}+x_{1,1} x_{2,0} x_{3,2}\right)+s_{10} \left(x_{1,2} x_{2,1} x_{3,1}+x_{1,1} x_{2,2} x_{3,1}+x_{1,1} x_{2,1} x_{3,2}\right)\\\nonumber &&+s_{11} \left(x_{1,2} x_{2,2} x_{3,0}+x_{1,2} x_{2,0} x_{3,2}+x_{1,0} x_{2,2} x_{3,2}\right)+s_9 \left(x_{1,2} x_{2,2} x_{3,1}+x_{1,2} x_{2,1} x_{3,2}+x_{1,1} x_{2,2} x_{3,2}\right) \\ \nonumber
\delta P_2 &=& s_8 \left(x_{1,1} x_{2,0} x_{3,0}+x_{1,0} x_{2,1} x_{3,0}+x_{1,0} x_{2,0} x_{3,1}\right)+s_6 \left(x_{1,1} x_{2,1} x_{3,0}+x_{1,1} x_{2,0} x_{3,1}+x_{1,0} x_{2,1} x_{3,1}\right)\\\nonumber &&+s_7 \left(x_{1,2} x_{2,0} x_{3,0}+x_{1,0} x_{2,2} x_{3,0}+x_{1,0} x_{2,0} x_{3,2}\right)+s_4 \left(x_{1,2} x_{2,1} x_{3,0}+x_{1,0} x_{2,2} x_{3,1}+x_{1,1} x_{2,0} x_{3,2}\right)\\\nonumber &&+s_5 \left(x_{1,1} x_{2,2} x_{3,0}+x_{1,2} x_{2,0} x_{3,1}+x_{1,0} x_{2,1} x_{3,2}\right)+s_3 \left(x_{1,2} x_{2,2} x_{3,0}+x_{1,2} x_{2,0} x_{3,2}+x_{1,0} x_{2,2} x_{3,2}\right)\\\nonumber &&+s_2 \left(x_{1,2} x_{2,2} x_{3,1}+x_{1,2} x_{2,1} x_{3,2}+x_{1,1} x_{2,2} x_{3,2}\right) \\ \nonumber
\delta P_3 &=& s_1 \left(x_{1,1} x_{2,0} x_{3,0}+x_{1,0} x_{2,1} x_{3,0}+x_{1,0} x_{2,0} x_{3,1}\right)
\end{eqnarray}
It should be noted that the apparent asymmetry between the $\delta P_i$ above is due, in part, to choice of conventions made in parameterizing the cokernels in question. For example, one could add an arbitrary combination of the $s$'s multiplied by the associated defining relation to one of these fluctuations and obtain an equally valid result.

With a good description of the complex structure fluctuations (\ref{csfluctbigone}) in hand we can now either compute the kernel in (\ref{ker1again}) or that in (\ref{ker2again}) with the dilaton taken to be fixed. As in the previous example, computing both maps simultaneously would require an understanding of how the various choices of maps are linked in the descriptions we are using. Here we will focus on (\ref{ker2again}) as this is the richer case, corresponding as it does (generically) to more than just one constraint.

The explicit expressions for the spaces (\ref{newnew2}) and (\ref{bigone}), for the complex structure given in (\ref{rancschoice}) and using the map defined using the quantities in (\ref{image}), can easily be computed explicitly and shown to depend upon the right number (16) of independent parameters. The expressions, while easily manipulated with a computer are too large to be included here and so we will content ourselves with discussing some of the results that can be obtained by using them.

The first observation that we can make in this more complicated example is that it very hard to find a flux that doesn't stabilize all of the complex structure moduli via (\ref{ker2again}) when the dilaton is held fixed. From a type IIB perspective this is completely expected, as the number of constraints being imposed, as determined by the target of the map, is the same as the number of degrees of freedom that we would wish to be fixed (16 in this case). Indeed, given that a general element of the cohomology class describing the map from $H^1(X,TX)$ in (\ref{ker2again}) can consistently be chosen as the flux, we would indeed expect generally to obtain a surjective map (and thus vanishing kernel). This is, however, somewhat different to the heterotic case. The heterotic example given in Section \ref{hetegs} is in fact of a similar nature. The target space in question is $h^2(X, \textnormal{End}_0(V))=17$ dimensional and indeed $17$ complex structure moduli are stabilized. More generally, however, the heterotic literature is full of examples where equation counting of this nature does not give the right answer. For example, in \cite{Anderson:2011ty} an explicit $SU(2)$ bundle is given for which the target space in the Atiyah calculation is $359$ dimensional but only $80$ complex structure moduli are fixed. It would be interesting to know if this break down in counting of constraints, or equivalently common non-genericity of the Atiyah map (\ref{map1}) in examples, is an artifact of the common bundle constructions that are used in the literature or a fairly generic feature of the holomorphic poly-stable bundles that can appear in heterotic compactifications.

Rather than simply picking a flux as an element of (\ref{bigone}) and computing the kernel of the map (\ref{ker2again}) with the dilaton fixed, one can perform the computation with a general map depending on $16$ unspecified parameters. The result of this computation is too large to include here for this more complicated example being a set of $81$ equations each with over $154$ terms. This set of equations is bilinear in the complex structure fluctuation parameters in (\ref{csfluctbigone}) and the $16$ parameters describing the possible fluxes. 
\begin{eqnarray} \label{IIBvariety}
\Lambda^{i \alpha}_I n_{\alpha}s_i
\end{eqnarray}
Here the index $I$ runs over the $81$ equations in the set.

It might be tempting to view these equations as analogous to the varieties obtained in (\ref{fulleq}) in the heterotic case and to perform primary decomposition and elimination on (\ref{IIBvariety}). The structure here is not the same however. The flux is not really allowed to vary in a continuous manner. It is, rather quantized, in a manner that we have largely ignored in this paper. This means that elimination is not the correct tool as this would tell us which complex structures could be accessed if the flux were allowed to vary smoothly and continuously.  

The root of this difference to the heterotic case is that in type IIB we have not provided a ``starting point" independent analysis of this system. Here we have simply picked an initial complex structure and have calculated a set of constraints linking the choice of flux to the allowed fluctuations in the complex structure moduli. This is somewhat different to the more global view on the moduli space that was afforded by the use of the extension construction in Section \ref{hetegs}. Thus the system of equations  (\ref{IIBvariety}) is not really analogous to (\ref{fulleq}) but rather to a set of equations that could be obtained involving possible choices of quantized cohomology class of field strength for the gauge bundle. It seems harder to obtain a true analog of (\ref{fulleq}) in the type IIB case as the analog of a bundle construction that is manifestly complex structure dependent does not seem obvious. One would need a description of the holomorphy of the flux that was global in moduli space, rather than just being analyzed by fluctuation around a single configuration as we have above.

As a final note, one might also worry in analyzing the equations of the form (\ref{IIBvariety}) in the type IIB string theory case that they are not valid for arbitrary values of the complex structure, but rather for infinitesimal values of the fluctuations $s_i$. In fact, this is not an issue as the equations are linear in the $s_i$. Therefore, for any solution that is found one can simply multiply all of the $s_i$ by an arbitrarily small number and obtain a valid solution where the fluctuations are small.

\section{Conclusions} \label{conc}

The question that was asked in the introduction to this paper was to what extent techniques of moduli identification that have been developed in examples of heterotic compactifications can be utilized in the case of compactifications of type IIB string theory. We have seen that these techniques can be applied to this case quite easily. Indeed in many situations the unstabilized moduli do correspond to kernels of maps between ordinary Dolbeault cohomology groups, with the maps being defined by the supergravity data of the compactification of interest. The main differences to the heterotic case arise once one attempts to compute these kernels in explicit examples.

We have seen that it is indeed possible to compute map kernels in examples based upon compactification on conformal CICY three-folds. Using a description of the cohomologies involved in terms of polynomials of ambient space homogeneous coordinates, the maps concerned can be carried out explicitly. Computing the relevant kernels is then straight forward and leads to an extremely explicit description of the unstabilized moduli, if any. In the case of complex structure moduli, for example, the unstabilized degrees of freedom can be described as unrestricted coefficients in the polynomial defining relations of the Calabi-Yau in the ambient product of projective spaces. 

There are, however, two important differences that arise to the heterotic case in performing this analysis. Firstly, a simplification occurs in that, in the polynomial description of the cohomologies in which the relevant maps live, any map can correspond to an allowed supergravity flux.  In the case of heterotic Calabi-Yau compactifications, where the relevant map is defined by the field strength of a non-abelian gauge field, the same is not true and there may be no appropriate bundle corresponding to a given choice of map. Secondly, a complication arises in the map structure we find in comparison to the heterotic case. The same field strength appears in two different maps in some examples, but in the form of its complex conjugate in one case. This leads to a difficulty in performing complete computations as identifying the complex conjugate elements in the two relevant spaces of maps is difficult in a description of the cohomologies based upon holomorphic polynomials. Nevertheless, this difficulty can be overcome in situations where the metric on complex structure moduli space is known in enough detail.

As in the heterotic case, the techniques presented here apply generally to manifolds described as complete intersections in simple ambient spaces. Thus they generalize easily to other popular constructions, such as that of hypersurfaces in toric varieties. It should be noted, however, that although the general analysis can be applied in many cases there still exists a dearth of suitable explicit examples to analyze in the case of compactifications that are not at least conformally Calabi-Yau.

Finally, it is interesting to ask whether the particular F-flat moduli spaces we have discussed here can be described in terms of the ordinary Dolbeault cohomology of a specific bundle with structure analogous to (for example) (\ref{atiyah1}) as seen in the heterotic setting. One might expect that this indeed could be the case given the importance of Courant algebroids in the study of generalized complex structures \cite{Gualtieri:2003dx,Hitchin:2004ut} and the structure of the ${\cal N}=1$ vacua being discussed here in terms of generalized Calabi-Yau manifolds \cite{Grana:2005jc,Grana:2005sn}. We leave such an investigation for future work.


\section*{Acknowledgments}

We would like to thank Lara Anderson and Eric Sharpe for valuable discussions. The work of J.G. and H.P. is supported in part by NSF grant PHY-1720321 and is part of the working group activities of the the 4-VA initiative ``A Synthesis of Two Approaches to String Phenomenology". 




\begin{thebibliography}{99}
\ifx\doiref\asklfhas\newcommand{\doiref}[2]{\href{http://dx.doi.org/#1}{#2}}\fi
\raggedright 
\ifx\arxivref\asklfhas\newcommand{\arxivref}[2]{\href{http://arxiv.org/abs/#1}{arXiv:#1}}\fi
\raggedright

\bibitem{Anderson:2010mh} 
  L.~B.~Anderson, J.~Gray, A.~Lukas and B.~Ovrut,
  ``Stabilizing the Complex Structure in Heterotic Calabi-Yau Vacua,''
  JHEP {\bf 1102}, 088 (2011)
  doi:10.1007/JHEP02(2011)088
  [arXiv:1010.0255 [hep-th]].
  
  \bibitem{Anderson:2011cza} 
  L.~B.~Anderson, J.~Gray, A.~Lukas and B.~Ovrut,
  ``Stabilizing All Geometric Moduli in Heterotic Calabi-Yau Vacua,''
  Phys.\ Rev.\ D {\bf 83}, 106011 (2011)
  doi:10.1103/PhysRevD.83.106011
  [arXiv:1102.0011 [hep-th]].
  
  \bibitem{Anderson:2011ty} 
  L.~B.~Anderson, J.~Gray, A.~Lukas and B.~Ovrut,
  ``The Atiyah Class and Complex Structure Stabilization in Heterotic Calabi-Yau Compactifications,''
  JHEP {\bf 1110}, 032 (2011)
  doi:10.1007/JHEP10(2011)032
  [arXiv:1107.5076 [hep-th]].

\bibitem{Anderson:2013qca} 
  L.~B.~Anderson, J.~Gray, A.~Lukas and B.~Ovrut,
  ``Vacuum Varieties, Holomorphic Bundles and Complex Structure Stabilization in Heterotic Theories,''
  JHEP {\bf 1307}, 017 (2013)
  doi:10.1007/JHEP07(2013)017
  [arXiv:1304.2704 [hep-th]].
  
  \bibitem{Anderson:2014xha} 
  L.~B.~Anderson, J.~Gray and E.~Sharpe,
  ``Algebroids, Heterotic Moduli Spaces and the Strominger System,''
  JHEP {\bf 1407}, 037 (2014)
  doi:10.1007/JHEP07(2014)037
  [arXiv:1402.1532 [hep-th]].

    \bibitem{Melnikov:2011ez} 
  I.~V.~Melnikov and E.~Sharpe,
  Phys.\ Lett.\ B {\bf 705}, 529 (2011)
  doi:10.1016/j.physletb.2011.10.055
  [arXiv:1110.1886 [hep-th]].
  
  \bibitem{delaOssa:2014cia} 
  X.~de la Ossa and E.~E.~Svanes,
  ``Holomorphic Bundles and the Moduli Space of N=1 Supersymmetric Heterotic Compactifications,''
  JHEP {\bf 1410}, 123 (2014)
  doi:10.1007/JHEP10(2014)123
  [arXiv:1402.1725 [hep-th]].


\bibitem{Goldstein:2002pg} 
  E.~Goldstein and S.~Prokushkin,
  ``Geometric model for complex nonKahler manifolds with SU(3) structure,''
  Commun.\ Math.\ Phys.\  {\bf 251}, 65 (2004)
  doi:10.1007/s00220-004-1167-7
  [hep-th/0212307].

\bibitem{Fu:2006vj} 
  J.~X.~Fu and S.~T.~Yau,
  ``The Theory of superstring with flux on non-Kahler manifolds and the complex Monge-Ampere equation,''
  J.\ Diff.\ Geom.\  {\bf 78}, no. 3, 369 (2008)
  [hep-th/0604063].

\bibitem{Fei:2015kua} 
  T.~Fei,
  ``A construction of non-Kähler Calabi?Yau manifolds and new solutions to the Strominger system,''
  Adv.\ Math.\  {\bf 302}, 529 (2016)
  doi:10.1016/j.aim.2016.07.023
  [arXiv:1507.00293 [math.DG]].

\bibitem{Fei:2017ctw} 
  T.~Fei, Z.~Huang and S.~Picard,
  ``A Construction of Infinitely Many Solutions to the Strominger System,''
  arXiv:1703.10067 [math.DG].

  
  

  \bibitem{Hubsch:1992nu} 
  T.~Hubsch,
  ``Calabi-Yau manifolds: A Bestiary for physicists,''


  
  \bibitem{DeWolfe:2002nn} 
  O.~DeWolfe and S.~B.~Giddings,
  ``Scales and hierarchies in warped compactifications and brane worlds,''
  Phys.\ Rev.\ D {\bf 67}, 066008 (2003)
  doi:10.1103/PhysRevD.67.066008
  [hep-th/0208123].

\bibitem{Grana:2003ek} 
  M.~Grana, T.~W.~Grimm, H.~Jockers and J.~Louis,
  ``Soft supersymmetry breaking in Calabi-Yau orientifolds with D-branes and fluxes,''
  Nucl.\ Phys.\ B {\bf 690}, 21 (2004)
  doi:10.1016/j.nuclphysb.2004.04.021
  [hep-th/0312232].

\bibitem{Grimm:2004uq} 
  T.~W.~Grimm and J.~Louis,
  ``The Effective action of N = 1 Calabi-Yau orientifolds,''
  Nucl.\ Phys.\ B {\bf 699}, 387 (2004)
  doi:10.1016/j.nuclphysb.2004.08.005
  [hep-th/0403067].

\bibitem{Grana:2004bg} 
  M.~Grana, R.~Minasian, M.~Petrini and A.~Tomasiello,
  ``Supersymmetric backgrounds from generalized Calabi-Yau manifolds,''
  JHEP {\bf 0408}, 046 (2004)
  doi:10.1088/1126-6708/2004/08/046
  [hep-th/0406137].

\bibitem{Behrndt:2005bv} 
  K.~Behrndt, M.~Cvetic and P.~Gao,
  ``General type IIB fluxes with SU(3) structures,''
  Nucl.\ Phys.\ B {\bf 721}, 287 (2005)
  doi:10.1016/j.nuclphysb.2005.05.020
  [hep-th/0502154].

\bibitem{Grana:2005sn} 
  M.~Grana, R.~Minasian, M.~Petrini and A.~Tomasiello,
  ``Generalized structures of N=1 vacua,''
  JHEP {\bf 0511}, 020 (2005)
  doi:10.1088/1126-6708/2005/11/020
  [hep-th/0505212].

\bibitem{Giddings:2005ff} 
  S.~B.~Giddings and A.~Maharana,
  ``Dynamics of warped compactifications and the shape of the warped landscape,''
  Phys.\ Rev.\ D {\bf 73}, 126003 (2006)
  doi:10.1103/PhysRevD.73.126003
  [hep-th/0507158].

\bibitem{Koerber:2006hh} 
  P.~Koerber and L.~Martucci,
  ``Deformations of calibrated D-branes in flux generalized complex manifolds,''
  JHEP {\bf 0612}, 062 (2006)
  doi:10.1088/1126-6708/2006/12/062
  [hep-th/0610044].

\bibitem{Tomasiello:2007zq} 
  A.~Tomasiello,
  ``Reformulating supersymmetry with a generalized Dolbeault operator,''
  JHEP {\bf 0802}, 010 (2008)
  doi:10.1088/1126-6708/2008/02/010
  [arXiv:0704.2613 [hep-th]].

\bibitem{Shiu:2008ry} 
  G.~Shiu, G.~Torroba, B.~Underwood and M.~R.~Douglas,
  ``Dynamics of Warped Flux Compactifications,''
  JHEP {\bf 0806}, 024 (2008)
  doi:10.1088/1126-6708/2008/06/024
  [arXiv:0803.3068 [hep-th]].

\bibitem{Douglas:2008jx} 
  M.~R.~Douglas and G.~Torroba,
  ``Kinetic terms in warped compactifications,''
  JHEP {\bf 0905}, 013 (2009)
  doi:10.1088/1126-6708/2009/05/013
  [arXiv:0805.3700 [hep-th]].

\bibitem{Frey:2008xw} 
  A.~R.~Frey, G.~Torroba, B.~Underwood and M.~R.~Douglas,
  ``The Universal Kahler Modulus in Warped Compactifications,''
  JHEP {\bf 0901}, 036 (2009)
  doi:10.1088/1126-6708/2009/01/036
  [arXiv:0810.5768 [hep-th]].

\bibitem{Marchesano:2008rg} 
  F.~Marchesano, P.~McGuirk and G.~Shiu,
  ``Open String Wavefunctions in Warped Compactifications,''
  JHEP {\bf 0904}, 095 (2009)
  doi:10.1088/1126-6708/2009/04/095
  [arXiv:0812.2247 [hep-th]].

\bibitem{Martucci:2009sf} 
  L.~Martucci,
  ``On moduli and effective theory of N=1 warped flux compactifications,''
  JHEP {\bf 0905}, 027 (2009)
  doi:10.1088/1126-6708/2009/05/027
  [arXiv:0902.4031 [hep-th]].

\bibitem{Chen:2009zi} 
  H.~Y.~Chen, Y.~Nakayama and G.~Shiu,
  ``On D3-brane Dynamics at Strong Warping,''
  Int.\ J.\ Mod.\ Phys.\ A {\bf 25}, 2493 (2010)
  doi:10.1142/S0217751X10048366
  [arXiv:0905.4463 [hep-th]].

\bibitem{Tseng:2009gr} 
  L.~S.~Tseng and S.~T.~Yau,
  ``Cohomology and Hodge Theory on Symplectic Manifolds. I.,''
  J.\ Diff.\ Geom.\  {\bf 91}, no. 3, 383 (2012)
  [arXiv:0909.5418 [math.SG]].

\bibitem{Underwood:2010pm} 
  B.~Underwood,
  ``A Breathing Mode for Warped Compactifications,''
  Class.\ Quant.\ Grav.\  {\bf 28}, 195013 (2011)
  doi:10.1088/0264-9381/28/19/195013
  [arXiv:1009.4200 [hep-th]].

\bibitem{Tseng:2010kt} 
  L.~S.~Tseng and S.~T.~Yau,
  ``Cohomology and Hodge Theory on Symplectic Manifolds. II,''
  J.\ Diff.\ Geom.\  {\bf 91}, no. 3, 417 (2012)
  [arXiv:1011.1250 [math.SG]].

\bibitem{Marchesano:2010bs} 
  F.~Marchesano, P.~McGuirk and G.~Shiu,
  ``Chiral matter wavefunctions in warped compactifications,''
  JHEP {\bf 1105}, 090 (2011)
  doi:10.1007/JHEP05(2011)090
  [arXiv:1012.2759 [hep-th]].

\bibitem{Grana:2011nb} 
  M.~Grana and F.~Orsi,
  ``N=1 vacua in Exceptional Generalized Geometry,''
  JHEP {\bf 1108}, 109 (2011)
  doi:10.1007/JHEP08(2011)109
  [arXiv:1105.4855 [hep-th]].

\bibitem{Coimbra:2011nw} 
  A.~Coimbra, C.~Strickland-Constable and D.~Waldram,
  ``Supergravity as Generalised Geometry I: Type II Theories,''
  JHEP {\bf 1111}, 091 (2011)
  doi:10.1007/JHEP11(2011)091
  [arXiv:1107.1733 [hep-th]].

\bibitem{Tseng:2011gv} 
  L.~S.~Tseng and S.~T.~Yau,
  ``Generalized Cohomologies and Supersymmetry,''
  Commun.\ Math.\ Phys.\  {\bf 326}, 875 (2014)
  doi:10.1007/s00220-014-1895-2
  [arXiv:1111.6968 [hep-th]].

\bibitem{Grimm:2012rg} 
  T.~W.~Grimm, D.~Klevers and M.~Poretschkin,
  ``Fluxes and Warping for Gauge Couplings in F-theory,''
  JHEP {\bf 1301}, 023 (2013)
  doi:10.1007/JHEP01(2013)023
  [arXiv:1202.0285 [hep-th]].

\bibitem{Coimbra:2012yy} 
  A.~Coimbra, C.~Strickland-Constable and D.~Waldram,
  ``Generalised Geometry and type II Supergravity,''
  Fortsch.\ Phys.\  {\bf 60}, 982 (2012)
  doi:10.1002/prop.201100096
  [arXiv:1202.3170 [hep-th]].

\bibitem{Frey:2013bha} 
  A.~R.~Frey and J.~Roberts,
  ``The Dimensional Reduction and K\"ahler Metric of Forms In Flux and Warping,''
  JHEP {\bf 1310}, 021 (2013)
  doi:10.1007/JHEP10(2013)021
  [arXiv:1308.0323 [hep-th]].

\bibitem{Grana:2014vva} 
  M.~Grana, J.~Louis, U.~Theis and D.~Waldram,
  ``Quantum Corrections in String Compactifications on SU(3) Structure Geometries,''
  JHEP {\bf 1501}, 057 (2015)
  doi:10.1007/JHEP01(2015)057
  [arXiv:1406.0958 [hep-th]].

\bibitem{Marchesano:2014iea} 
  F.~Marchesano, D.~Regalado and G.~Zoccarato,
  ``On D-brane moduli stabilisation,''
  JHEP {\bf 1411}, 097 (2014)
  doi:10.1007/JHEP11(2014)097
  [arXiv:1410.0209 [hep-th]].

\bibitem{Martucci:2014ska} 
  L.~Martucci,
  ``Warping the Kähler potential of F-theory/IIB flux compactifications,''
  JHEP {\bf 1503}, 067 (2015)
  doi:10.1007/JHEP03(2015)067
  [arXiv:1411.2623 [hep-th]].

\bibitem{Coimbra:2014uxa} 
  A.~Coimbra, C.~Strickland-Constable and D.~Waldram,
  ``Supersymmetric Backgrounds and Generalised Special Holonomy,''
  Class.\ Quant.\ Grav.\  {\bf 33}, no. 12, 125026 (2016)
  doi:10.1088/0264-9381/33/12/125026
  [arXiv:1411.5721 [hep-th]].

\bibitem{Grimm:2014efa} 
  T.~W.~Grimm, T.~G.~Pugh and M.~Weissenbacher,
  ``The effective action of warped M-theory reductions with higher derivative terms ? part I,''
  JHEP {\bf 1601}, 142 (2016)
  doi:10.1007/JHEP01(2016)142
  [arXiv:1412.5073 [hep-th]].

\bibitem{Coimbra:2015nha} 
  A.~Coimbra and C.~Strickland-Constable,
  ``Generalised Structures for $\mathcal{N}=1$ AdS Backgrounds,''
  JHEP {\bf 1611}, 092 (2016)
  doi:10.1007/JHEP11(2016)092
  [arXiv:1504.02465 [hep-th]].

\bibitem{Grimm:2015mua} 
  T.~W.~Grimm, T.~G.~Pugh and M.~Weissenbacher,
  ``The effective action of warped M-theory reductions with higher-derivative terms - Part II,''
  JHEP {\bf 1512}, 117 (2015)
  doi:10.1007/JHEP12(2015)117
  [arXiv:1507.00343 [hep-th]].

\bibitem{Carta:2016ynn} 
  F.~Carta, F.~Marchesano, W.~Staessens and G.~Zoccarato,
  ``Open string multi-branched and Kähler potentials,''
  JHEP {\bf 1609}, 062 (2016)
  doi:10.1007/JHEP09(2016)062
  [arXiv:1606.00508 [hep-th]].

\bibitem{Cownden:2016hpf} 
  B.~Cownden, A.~R.~Frey, M.~C.~D.~Marsh and B.~Underwood,
  ``Dimensional Reduction for D3-brane Moduli,''
  JHEP {\bf 1612}, 139 (2016)
  doi:10.1007/JHEP12(2016)139
  [arXiv:1609.05904 [hep-th]].

\bibitem{Martucci:2016pzt} 
  L.~Martucci,
  ``Warped Kähler potentials and fluxes,''
  JHEP {\bf 1701}, 056 (2017)
  doi:10.1007/JHEP01(2017)056
  [arXiv:1610.02403 [hep-th]].

\bibitem{Sethi:2017phn} 
  S.~Sethi,
  ``Supersymmetry Breaking by Fluxes,''
  arXiv:1709.03554 [hep-th].


  
  
   \bibitem{Witten:1985bz} 
  E.~Witten,
  ``New Issues in Manifolds of SU(3) Holonomy,''
  Nucl.\ Phys.\ B {\bf 268}, 79 (1986).
  doi:10.1016/0550-3213(86)90202-6

\bibitem{Anderson:2009sw} 
  L.~B.~Anderson, J.~Gray, A.~Lukas and B.~Ovrut,
  ``The Edge Of Supersymmetry: Stability Walls in Heterotic Theory,''
  Phys.\ Lett.\ B {\bf 677}, 190 (2009)
  doi:10.1016/j.physletb.2009.05.025
  [arXiv:0903.5088 [hep-th]].

\bibitem{Gukov:1999ya} 
  S.~Gukov, C.~Vafa and E.~Witten,
  ``CFT's from Calabi-Yau four folds,''
  Nucl.\ Phys.\ B {\bf 584}, 69 (2000)
  Erratum: [Nucl.\ Phys.\ B {\bf 608}, 477 (2001)]
  doi:10.1016/S0550-3213(01)00289-9, 10.1016/S0550-3213(00)00373-4
  [hep-th/9906070].
      
  \bibitem{Grana:2005jc} 
  M.~Grana,
  ``Flux compactifications in string theory: A Comprehensive review,''
  Phys.\ Rept.\  {\bf 423}, 91 (2006)
  doi:10.1016/j.physrep.2005.10.008
  [hep-th/0509003].
      
  \bibitem{Strominger:1986uh} 
  A.~Strominger,
  ``Superstrings with Torsion,''
  Nucl.\ Phys.\ B {\bf 274}, 253 (1986).
  doi:10.1016/0550-3213(86)90286-5
        
  \bibitem{Koerber:2007xk} 
  P.~Koerber and L.~Martucci,
  ``From ten to four and back again: How to generalize the geometry,''
  JHEP {\bf 0708}, 059 (2007)
  doi:10.1088/1126-6708/2007/08/059
  [arXiv:0707.1038 [hep-th]].
  
  
  
 
 
\bibitem{Becker:1996gj} 
  K.~Becker and M.~Becker,
  ``M theory on eight manifolds,''
  Nucl.\ Phys.\ B {\bf 477}, 155 (1996)
  doi:10.1016/0550-3213(96)00367-7
  [hep-th/9605053].

\bibitem{Dasgupta:1999ss} 
  K.~Dasgupta, G.~Rajesh and S.~Sethi,
  ``M theory, orientifolds and G - flux,''
  JHEP {\bf 9908}, 023 (1999)
  doi:10.1088/1126-6708/1999/08/023
  [hep-th/9908088].

\bibitem{Greene:2000gh} 
  B.~R.~Greene, K.~Schalm and G.~Shiu,
  ``Warped compactifications in M and F theory,''
  Nucl.\ Phys.\ B {\bf 584}, 480 (2000)
  doi:10.1016/S0550-3213(00)00400-4
  [hep-th/0004103].

\bibitem{Giddings:2001yu} 
  S.~B.~Giddings, S.~Kachru and J.~Polchinski,
  ``Hierarchies from fluxes in string compactifications,''
  Phys.\ Rev.\ D {\bf 66}, 106006 (2002)
  doi:10.1103/PhysRevD.66.106006
  [hep-th/0105097].

 
 
 \bibitem{Yau:1986gu} 
  S.~T.~Yau,
  ``Compact Three-dimensional Kahler Manifolds With Zero Ricci Curvature,''
  In *Argonne/chicago 1985, Proceedings, Anomalies, Geometry, Topology*, 395-406

\bibitem{Hubsch:1986ny} 
  T.~Hubsch,
  ``Calabi-yau Manifolds: Motivations and Constructions,''
  Commun.\ Math.\ Phys.\  {\bf 108}, 291 (1987).
  doi:10.1007/BF01210616

\bibitem{Green:1986ck} 
  P.~Green and T.~Hubsch,
  ``Calabi-yau Manifolds as Complete Intersections in Products of Complex Projective Spaces,''
  Commun.\ Math.\ Phys.\  {\bf 109}, 99 (1987).
  doi:10.1007/BF01205673

\bibitem{Candelas:1987kf} 
  P.~Candelas, A.~M.~Dale, C.~A.~Lutken and R.~Schimmrigk,
  ``Complete Intersection Calabi-Yau Manifolds,''
  Nucl.\ Phys.\ B {\bf 298}, 493 (1988).
  doi:10.1016/0550-3213(88)90352-5

\bibitem{Candelas:1987du} 
  P.~Candelas, C.~A.~Lutken and R.~Schimmrigk,
  ``Complete Intersection Calabi-yau Manifolds. 2. Three Generation Manifolds,''
  Nucl.\ Phys.\ B {\bf 306}, 113 (1988).
  doi:10.1016/0550-3213(88)90173-3
 
 
 \bibitem{Anderson:2015iia} 
  L.~B.~Anderson, F.~Apruzzi, X.~Gao, J.~Gray and S.~J.~Lee,
  ``A new construction of Calabi-Yau manifolds: Generalized CICYs,''
  Nucl.\ Phys.\ B {\bf 906}, 441 (2016)
  doi:10.1016/j.nuclphysb.2016.03.016
  [arXiv:1507.03235 [hep-th]].
  
 \bibitem{Berglund:2016yqo} 
  P.~Berglund and T.~Hubsch,
  ``On Calabi-Yau generalized complete intersections from Hirzebruch varieties and novel K3-fibrations,''
  arXiv:1606.07420 [hep-th].

\bibitem{Berglund:2016nvh} 
  P.~Berglund and T.~Hubsch,
  ``A Generalized Construction of Calabi-Yau Models and Mirror Symmetry,''
  SciPost Phys.\  {\bf 4}, 009 (2018)
  doi:10.21468/SciPostPhys.4.2.009
  [arXiv:1611.10300 [hep-th]].


\bibitem{Garbagnati:2017rtb} 
  A.~Garbagnati and B.~van Geemen,
  ``A remark on generalized complete intersections,''
  Nucl.\ Phys.\ B {\bf 925}, 135 (2017)
  doi:10.1016/j.nuclphysb.2017.10.006
  [arXiv:1708.00517 [math.AG]].
  
  
  \bibitem{cicy} L.B.~Anderson, J.~Gray, Y.-H.~He, S.-J.~Lee and A.~Lukas 'The Cicy Package', based on methods described in arXiv:0911.1569, arXiv:0911.0865, arXiv:0805.2875, hep-th/0703249, hep-th/0702210.
  
\bibitem{Candelas:2008wb} 
  P.~Candelas and R.~Davies,
  ``New Calabi-Yau Manifolds with Small Hodge Numbers,''
  Fortsch.\ Phys.\  {\bf 58}, 383 (2010)
  doi:10.1002/prop.200900105
  [arXiv:0809.4681 [hep-th]].

\bibitem{Braun:2010vc} 
  V.~Braun,
  ``On Free Quotients of Complete Intersection Calabi-Yau Manifolds,''
  JHEP {\bf 1104}, 005 (2011)
  doi:10.1007/JHEP04(2011)005
  [arXiv:1003.3235 [hep-th]].

\bibitem{Candelas:2010ve} 
  P.~Candelas and A.~Constantin,
  ``Completing the Web of $Z_3$ - Quotients of Complete Intersection Calabi-Yau Manifolds,''
  Fortsch.\ Phys.\  {\bf 60}, 345 (2012)
  doi:10.1002/prop.201200044
  [arXiv:1010.1878 [hep-th]].

\bibitem{Candelas:2015amz} 
  P.~Candelas, A.~Constantin and C.~Mishra,
  ``Hodge Numbers for CICYs with Symmetries of Order Divisible by 4,''
  Fortsch.\ Phys.\  {\bf 64}, no. 6-7, 463 (2016)
  doi:10.1002/prop.201600005
  [arXiv:1511.01103 [hep-th]].

\bibitem{Candelas:2016fdy} 
  P.~Candelas, A.~Constantin and C.~Mishra,
  ``Calabi-Yau Threefolds With Small Hodge Numbers,''
  arXiv:1602.06303 [hep-th].

\bibitem{Constantin:2016xlj} 
  A.~Constantin, J.~Gray and A.~Lukas,
  ``Hodge Numbers for All CICY Quotients,''
  JHEP {\bf 1701}, 001 (2017)
  doi:10.1007/JHEP01(2017)001
  [arXiv:1607.01830 [hep-th]].

\bibitem{Candelas:1990pi} 
  P.~Candelas and X.~de la Ossa,
  ``Moduli Space of {Calabi-Yau} Manifolds,''
  Nucl.\ Phys.\ B {\bf 355}, 455 (1991).
  doi:10.1016/0550-3213(91)90122-E

\bibitem{Candelas:1990rm} 
  P.~Candelas, X.~C.~De La Ossa, P.~S.~Green and L.~Parkes,
  ``A Pair of Calabi-Yau manifolds as an exactly soluble superconformal theory,''
  Nucl.\ Phys.\ B {\bf 359}, 21 (1991)
  [AMS/IP Stud.\ Adv.\ Math.\  {\bf 9}, 31 (1998)].
  doi:10.1016/0550-3213(91)90292-6

\bibitem{Hitchin:2004ut} 
  N.~Hitchin,
  ``Generalized Calabi-Yau manifolds,''
  Quart.\ J.\ Math.\  {\bf 54}, 281 (2003)
  doi:10.1093/qjmath/54.3.281
  [math/0209099 [math-dg]].

\bibitem{Gualtieri:2003dx} 
  M.~Gualtieri,
  ``Generalized complex geometry,''
  math/0401221 [math-dg].


 \end{thebibliography}
\end{document}